\documentclass[fleqn,usenatbib]{mnras}

\usepackage{amssymb}

\usepackage{newtxtext,newtxmath}

\usepackage[T1]{fontenc}

\DeclareRobustCommand{\VAN}[3]{#2}
\let\VANthebibliography\thebibliography
\def\thebibliography{\DeclareRobustCommand{\VAN}[3]{##3}\VANthebibliography}

\usepackage{graphicx}	
\usepackage{amsmath}	
\usepackage{amssymb}	
\usepackage{physics}
\usepackage{fancyvrb}
\usepackage[dvipsnames]{xcolor}
\usepackage{booktabs}
\usepackage{comment}

\RecustomVerbatimCommand{\VerbatimInput}{VerbatimInput}%
{fontsize=\footnotesize,
 frame=lines,  
 framesep=2em, 
 rulecolor=\color{Gray},
 label=\fbox{\color{Black}inlist\_project},
 labelposition=topline,
 commandchars=\|\(\), 
 commentchar=*        
}

\title[Instabilities of massive stars]{Viscous and centrifugal instabilities of massive stars}

\author[Y. Shi \& J. Fuller]{
Yanlong Shi$^{1}$\thanks{E-mail: yanlong@caltech.edu} and
Jim Fuller$^{1}$\thanks{E-mail: jfuller@caltech.edu}
\\
$^1$TAPIR, MC 350-17, California Institute of Technology, Pasadena, CA 91125
}

\date{Accepted XXX. Received YYY; in original form ZZZ}

\pubyear{2021}

\newif\ifshowcomments
\showcommentstrue

\newif\ifshowcomments
\showcommentstrue

\begin{document}
\label{firstpage}
\pagerange{\pageref{firstpage}--\pageref{lastpage}}
\maketitle

\begin{abstract}

Massive stars exhibit a variety of instabilities, many of which are poorly understood. We explore instabilities induced by centrifugal forces and angular momentum transport in massive rotating stars. First, we derive and numerically solve linearized oscillation equations for adiabatic radial modes in polytropic stellar models. In the presence of differential rotation, we show that centrifugal and Coriolis forces combined with viscous angular momentum transport can excite stellar pulsation modes, under both low- or high-viscosity conditions. In the low-viscosity limit, which is common in real stars, we demonstrate how to compute mode growth/damping rates via a work integral. Finally, we build realistic rotating $30\,M_\odot$ star models  and show that overstable (growing) radial modes are predicted to exist for most of the star's life, in the absence of non-adiabatic effects. Peak growth rates are predicted to occur while the star is crossing the Hertzsprung-Russell gap, though non-adiabatic damping may dominate over viscous driving, depending on the effective viscosity produced by convective and/or magnetic torques. Viscous instability could be a new mechanism to drive massive star pulsations and is possibly related to instabilities of luminous blue variable stars.

\end{abstract}




\section{Introduction}

Massive stars can lose a significant amount of mass during their evolution, and the physical mechanisms behind the mass loss are not well understood.  One well studied mechanism is line-driven winds (e.g., \citealt{vink:21}), which are caused primarily by scattering between photons and iron-group elements. Another well studied mechanism is envelope stripping by a binary companion, which occurs in a large fraction of binary stars \citep{sana:12}. \cite{smithrev:14} presents a comprehensive review of theories and observations of massive star mass loss.

However, it is becoming increasingly clear that some stars such as luminous blue stars (LBVs, or S Doradus variables) lose their mass through outbursts or eruptions (see \citealt{smith:17}, \citealt{davidson:20} for recent reviews). While it is generally agreed that radiation pressure in these near-Eddington (or super-Eddington) stars is an important factor for their mass loss (e.g., \citealt{jiang:18}), the details of the outburst mechanisms are not well understood. Moreover, supernovae observations have made it abundantly clear that a significant fraction of massive stars undergo outbursts or enhanced mass loss in the final years of their lives (see e.g., \citealt{wu:21} for a list of examples). The mechanisms underlying LBV eruptions and pre-supernova outbursts are highly debated, motivating continued studies of massive star instabilities.

One possible mechanism for variability and mass loss in LBVs are instabilities related to rapid stellar rotation. Indeed, \cite{groh:06,groh:09} have shown the some LBVs do appear to rotate very rapidly, near their {critical} rotation rate of $\Omega_{\rm crit} \approx \sqrt{GM_\star/R^3}$ (where $M_\star$ and $R$ are mass and radius of the star). \cite{ZhaoFuller2020} showed that efficient angular momentum transport can cause massive stars evolving off the main sequence (and through the LBV instability region) to have surface rotation rates that approach {the critical rate} (see also \citealt{langer:98,ekstrom:08,HastingsWang2020}), driving centrifugally enhanced mass loss. However, it is not clear whether the mass ejection mechanism would be centrifugally enhanced winds (e.g., \citealt{gagnier:19}), loss of mass into a decretion disk, or whether mass loss occurs via rotationally enhanced instabilities.

\cite{ZhaoFuller2020} demonstrated that the stellar envelopes can be (in a local sense) unstable in the presence of rapid rotation, especially the layers of the star near the iron opacity bump where the adiabatic index of the star is small. However, their simple calculation employed a local approximation (i.e., they did not perform a global stability calculation) and only considered the parameterized limits of zero angular momentum (AM) transport or instantaneous AM transport over the pulsation cycle. Efficient AM transport increases instability because it allows mass elements perturbed outward to gain AM from inner layers, further accelerating the outer layers via the centrifugal force. However, it is not clear whether a global centrifugal instability actually occurs in realistic stars with a finite angular momentum transport time scale between different stellar layers.

In this paper, we examine the possibility of rotationally driven instabilities in massive stars by computing radial oscillation mode frequencies for realistic stellar models. Instability could come in two flavors. The first is a centrifugal instability (i.e., imaginary mode frequency) due to efficient AM transport as suggested by \cite{ZhaoFuller2020}. The second is a viscous overstability (i.e., complex mode frequency) sourced by differential rotation of the star and harnessed by an effective AM viscosity. Section \ref{sec:linear} describes our equation and computational method, Section \ref{sec:polytropic} applies these calculations to simple polytropic models to illustrate the concepts, and Section \ref{sec:mesa} applies the calculations to realistic stellar models. We discuss observational implications and physical uncertainties in Section \ref{sec:discussion}, and we conclude in Section \ref{sec:conclusions}.

\section{Rotational instability}
\label{sec:linear}


In this section, we derive perturbation (oscillation) equations from the equations of motion in a rotating star, including centrifugal forces and viscous angular momentum transport. Then we discuss behaviors of oscillation modes under low- or high-viscosity limits. Finally, we derive an approximate expression for mode growth/decay rates in the low-viscosity limit, and discuss conditions for the instability to arise.

\subsection{Equation of Motion}

We assume a star locally rotating at an angular frequency of $\Omega(r)$ which is a function of radius $r$ with rotation axis in the $z$ direction. We assume $\Omega$ is well below the {critical} limit such that quantities like density and pressure are still functions of only radius (i.e., latitudinal variations are ignored), and gravity remains in the radial direction. {For each radius, we adopt a reference frame that is locally co-rotating with the star at that point.} For a mass element located at spherical coordinate $(r,\theta,\phi)$, the equation of motion in a co-rotating frame is
\footnote{Starting from the inertial frame generates the same equations of motion. The inertial frame velocity $\vec{v}_{\rm I}$ and rotating frame velocity $\vec{v}_{\rm R}$ are related by 
    \begin{align}
        \frac{\dd \vec{v}_{\rm I}}{\dd t} = \left(\frac{\dd}{\dd t} + \vec{\Omega} \times  \right) \vec{v}_{\rm R}. \nonumber
    \end{align}
{The rotation rate $\Omega$ can be expressed as $\Omega =\dot{\phi}_R$, where $\phi_R$ is the azimuthal coordinate in the inertial frame.} Hence, the fluid element position varies as
    \begin{align}
        \frac{\dd^2 \vec{r}_{\rm I}}{\dd t^2} = \left(\frac{\dd}{\dd t} + \vec{\Omega} \times \right)^2 \vec{r}_{\rm R} = \ddot{\vec{r}}_{\rm R} {+\dot{\vec{\Omega}}\times \vec{r}_{\rm R}} + 2 \vec{\Omega} \times \dot{\vec{r}}_{\rm R} + \vec{\Omega} \times (\vec{\Omega} \times \vec{r}_{\rm R}). \nonumber
    \end{align}
Therefore, working with total (Lagrangian) time derivatives in the rotating frame yields equation \ref{equ:euler}, and there are no extra terms that arise due to differential rotation. 
}
\begin{align}
    \ddot{\vec{r}} {+\dot{\vec{\Omega}}\times \vec{r}} = -\frac{\nabla p}{\rho} + \vec{f}_{\rm grav} + \vec{f}_{\rm cor}+ f_{\rm cen} \left(\sin \theta \hat{r} + \cos \theta \hat{\theta} \right) \nonumber \\ + f_{\rm vis} \hat{\phi}+{\vec{f}_{\rm ext}}.\label{equ:euler}
\end{align}
Here $\ddot{\vec{r}} = \dd^2 \vec{r}/\dd t^2$ and $\dd/\dd t$ is the total derivative (or material derivative); $p(r)$ is the pressure and $\rho(r)$ is the density; $\vec{f}_{\rm grav} = -GM \hat{r}/r^2$ where $M(r)$ is the total mass enclosed inside radius $r$; $\vec{f}_{\rm cor}=- 2 \vec{\Omega} \times \vec{v}$ is the Coriolis force where $\vec{v}=\dot{\vec{r}}$ is the velocity. The centrifugal force is decomposed into two directions, with amplitude $f_{\rm cen} = \Omega ^2 r\sin \theta$. The $\dot{\vec{\Omega}}\times \vec{r}$ term is the Euler force, {which vanishes under equilibrium, but is non-zero after perturbation.}

Above, $f_{\rm vis}$ is the viscous force due to differential rotation:
\begin{align}
\label{equ:fvis}
    f_{\rm vis} = \frac{1}{\rho r^3 } \frac{\partial}{\partial r} \left(\rho r^4 \nu \frac{\partial \Omega}{\partial r} \sin \theta \right).
\end{align}
Although AM transport (e.g., via magnetic or turbulent stresses) need not act viscously, we model it with an effective viscosity $\nu$, and emphasize that $\nu$ can be much greater than the microscopic viscosity. With this form of AM transport, the total torque on the star is $\dot{J} = \int f_\phi r \sin \theta \dd M \propto \rho r^4 \nu \partial \Omega/\partial r \rvert_0^{R}=0$, which means the total AM is conserved. 

In the equilibrium state, we have $\ddot{\vec{r}} = \dot{\vec{r}}=0$, thus the Coriolis force is zero, {and the viscous force should be canceled out by some external force, which is the $\vec{f}_{\rm ext}$ term in  Eq. \eqref{equ:euler}}. Averaging over latitudinal coordinate, we obtain radial hydrostatic equilibrium
{
\begin{align}
    \frac{1}{\rho}\frac{\dd p}{\dd r} + \frac{GM}{r^2} -\frac{2}{3}\Omega^2 r = 0 \, ,
\end{align}
}with the average centrifugal force in the radial direction $f_{\rm cen} = 2\Omega^2 r/3$.
The centrifugal force can be a source of instability in massive rotating stars \cite[]{ZhaoFuller2020}, which we examine below.
Since we will only consider radial modes in this paper, we spherically average the equations over the $\theta$ direction, which means an effective multiplier of $\sqrt{2/3}$ in the rotational rate. For simplicity we absorb the multiplier and redefine $\Omega$ as $\sqrt{3/2} \Omega$ such that the outward centrifugal force is $\Omega^2 r$. We also define the dimensionless variables based on the {critical} rotational rate of the star. For a star with mass $M_\star$ and radius $R$, the {critical} rotational rate is defined as $\Omega_{\rm crit}=\sqrt{G M_\star/R^3}$. Dimensionless variables are denoted with a tilde, e.g., $\tilde{\Omega}=\Omega/\Omega_{\rm crit}$, $\tilde{\omega}=\omega/\Omega_{\rm crit}$, and $\tilde{\nu} = \nu/(\Omega_{\rm crit} R^2)$.

\subsection{Lagrangian Perturbations}

We now introduce linear perturbations to the equilibrium structure to calculate oscillation modes \cite[see, e.g.,][]{Dziembowski1971}. We consider radial modes with no angular dependence. For each mode, the perturbation can be decomposed into products of a radial eigenfunction, and a corresponding oscillatory part $e^{-i\omega t}$, where $\omega$ is the oscillation frequency. Naturally, if the imaginary part of $\omega$ is positive, the mode will grow exponentially.
We use a Lagrangian perturbation formalism,
{such that shells of the star are displaced as $r \to r + \delta r(r,t)$, $\Omega \to \Omega+\delta \Omega(r, t)$, $p\to p+ \delta p(r,t)$, and $\rho\to \rho+ \delta \rho(r,t)$.
Each small perturbation is a function of time $t$, and the initial (unperturbed) radius $r$}.

The first basic relation is mass conservation, which generates the continuity equation $\dd M = 4 \pi r^2 \rho \dd r$. The mass inside the shell $(r, r+\dd r)$ is conserved under the small perturbation, thus we have 
\begin{align}
    \frac{\dd }{\dd r} \left(\frac{\delta r}{r} \right) = -\frac{1}{r} \left(3 \frac{\delta r}{r} + \frac{1}{\Gamma_1}\frac{\delta p}{p}\right).\label{equ:mass-conservation}
\end{align}
Here $\Gamma_1$ is the adiabatic index,
and we assume an adiabatic equation of state such that $\delta \rho /\rho = (\delta p/p)/\Gamma_1$.

We then perturb the momentum equation (Eq. \ref{equ:euler}). In the radial direction, we have
\begin{align}
     \delta \ddot{ r}  & = -\frac{p}{\rho} \frac{\dd }{\dd r} \left(\frac{\delta p}{p}\right)    - \frac{1}{\rho}\frac{\dd p}{\dd r} \left(\frac{\delta p}{p}+2\frac{\delta r}{r} \right)  + 2\frac{GM}{r^2} \frac{\delta r}{r} \nonumber
     \\ & + \Omega^2 r \left(2\frac{\delta \Omega}{\Omega}+\frac{\delta r}{r}\right) \label{equ:euler-radial}.
\end{align}

The $\phi$-component of the momentum equation is a little more tricky. In order for the background state to be in equilibrium, we imagine a force $f_{\rm ext}$ that opposes the viscous force $f_{\rm vis}$ such that the background state is in equilibrium and { $\ddot{\phi} = d \Omega/dt = 0$.} We imagine $f_{\rm ext}$ is not affected by a linear perturbation. Physically, this is similar to stellar evolutionary effects (e.g., magnetic braking or spin-up via contraction) that maintain differential rotation in spite of viscous restoring forces.
Since the rotation rate is $\Omega = \dot{\phi}$, we have $\delta \dot{\phi} = \delta \Omega$ and $\delta \ddot{\phi} = \delta \dot{ \Omega}$. {We are using a frame that is always co-rotating with the shell, so the $\phi$-coordinates do not change throughout the perturbation cycle and the $\ddot{\vec{r}}$ term of equation \ref{equ:euler} is zero in the $\phi$-direction, but the Euler term is non-zero.}
We also need to consider perturbations in the Coriolis force and viscous force. The full linear perturbation gives
\begin{align}
    \frac{\delta \dot{\Omega}}{\Omega} + 2 \frac{\delta \dot{r}}{r} & = \frac{q \nu }{r^2} r\frac{\dd }{\dd r} \left[ \frac{2}{\Gamma_1}\frac{\delta p}{p} +6 \frac{\delta r}{r} + \frac{\delta \nu}{\nu} + \frac{\delta \Omega}{\Omega}  \right.\nonumber \\
    & + \left. \frac{1}{q} r\frac{\dd }{\dd r} \left(\frac{\delta \Omega}{\Omega} \right) \right]. \label{equ:euler-horizontal}
\end{align}
Here 
\begin{equation}
    q = \frac{ \dd \ln \Omega}{\dd \ln r} 
\end{equation}
is the background shear. We expect $q<0$ since stellar cores typically rotate faster than their surfaces.
Time derivatives in the above equations can be replaced with a factor of $-i\omega$, so we treat the eigenfunctions $\delta r/r$, $\delta p /p$, and $\delta \Omega / \Omega$ as functions of $r$ only.

{The same equations can also be derived from the inertial frame. \cite{unno:89} studied a similar problem using Eulerian perturbations applied to an inertial frame. In Appendix \ref{sec:eulerian_perturbations}, we translate the Eulerian perturbation equations of \cite{unno:89} into Lagrangian ones as we defined in this article. We show that Eqs. \eqref{equ:euler-radial} and \eqref{equ:euler-horizontal} are fully recovered, hence the two methods yield identical results.
}

As an approximation, \cite{ZhaoFuller2020} assumed a linear relation between the perturbations in rotational rate and radius,
\begin{align}
    \frac{\delta \Omega}{\Omega} = f_\Omega \frac{\delta r}{r}.
\end{align}
AM conservation (i.e., no viscous torques) would imply $f_\Omega = -2$, while $f_\Omega \sim 0$ would be expected for very rapid AM transport. If $f_\Omega$ is assumed to be constant, then Eq. \eqref{equ:mass-conservation} and Eq. \eqref{equ:euler-radial} can be solved without considering Eq. \eqref{equ:euler-horizontal}.
If we further decompose the perturbations into $e^{-i(\omega t - kr)}$, where $k$ is the wave number, we find that in  the long-wavelength limit ($k\to 0$), Eqs. \eqref{equ:mass-conservation} and \eqref{equ:euler-radial} combine to yield
\begin{align}
    r\omega^2 \approx  (3 \Gamma_1 -4) f_{\rm grav} - (2f_{\Omega} + 3\Gamma_1 -1) f_{\rm cen},
\end{align}
which is the same dispersion relation given in \cite{ZhaoFuller2020}. Note that large viscous torques ($f_\Omega \sim 0)$ are more destabilizing than zero viscous torques ($f_\Omega =-2$).

In what follows, we neglect the perturbation in viscosity, $\delta \nu$, though that term should be included in more detailed calculations based on a specific form of viscosity (e.g., magnetic or convective viscous torques). To solve the equations above, we first redefine some variables and get 4 first-order linear equations. The 4 dimensionless variables are
\begin{align}
    \begin{split}
        y_1 & = \frac{\delta r}{r}, \qquad
        y_2  = \frac{\delta p}{p}, \\
        y_3 & = \frac{\delta \Omega}{\Omega}, \qquad
        y_4  = \frac{2}{\Gamma_1}\frac{\delta p}{p} +6 \frac{\delta r}{r} + \frac{\delta \nu}{\nu} + \frac{\delta \Omega}{\Omega}  + \frac{1}{q} r\frac{\dd }{\dd r} \left(\frac{\delta \Omega}{\Omega} \right).\label{equ:y-defs}
    \end{split}
\end{align}
Then Eqs. (\ref{equ:mass-conservation}--\ref{equ:euler-horizontal}) can be written as
\begin{align}
    r \frac{\dd y_1}{\dd r} & = - \left(3 y_1 + \frac{1}{\Gamma_1} y_2\right), \label{equ:4-parameter-equations-y1}\\
    r \frac{\dd y_2}{\dd r} & = \frac{GM\rho}{r p} \left[\left(4+ \frac{r^3 \omega^2}{GM} - \frac{r^3 \Omega^2}{GM} \right)y_1 + \left(1-\frac{r^3 \Omega^2}{GM}\right) y_2\right. \nonumber \\
    & +\left.2\frac{r^3 \Omega^2}{GM} y_3 \right], \label{equ:4-parameter-equations-y2}\\
    r \frac{\dd y_3}{\dd r} & = q \left(-6 y_1 - \frac{2}{\Gamma_1} y_2 -y_3 + y_4 \right), \label{equ:4-parameter-equations-y3} \\
    r \frac{\dd y_4}{\dd r} & = -i\frac{\omega r^2}{q \nu} \left(2 y_1 + y_3 \right).\label{equ:4-parameter-equations-y4}
\end{align}
With four equations, four boundary conditions must be imposed based on physical considerations. We enumerate them below.
\begin{enumerate}
    \item At $r=0$, {the left hand side of Eq. \eqref{equ:4-parameter-equations-y1} is zero, so the right hand side must also be zero, requiring}
    \begin{align}
        3 y_1 + \frac{1}{\Gamma_1} y_2 = 0.\label{equ:bc1}
    \end{align}
    \item At $r=0$, {a non-diverging value of $\ddot{\phi}$ due to the viscous term (Eq. \ref{equ:fvis}) requires that $q \propto r^2$ (to lowest order in $r$) near the origin. The left hand side of Eq. \eqref{equ:4-parameter-equations-y4} is zero, which means a vanishing right hand side requires}:
    \begin{align}
        2y_1+y_3 = 0.\label{equ:bc2}
    \end{align}
    \item At $r=R$ (or at the surface), we see the term $GM\rho/(rp)$ diverges in Eq. \eqref{equ:4-parameter-equations-y2}. To reconcile the divergence, we must have
    \begin{align}
        (4 + \tilde{\omega}^2 -\tilde{\Omega}^2 ) y_1 + (1 - \tilde{\Omega}^2) y_2 + 2 \tilde{\Omega}^2 y_3 & = 0.\label{equ:bc3}
    \end{align}
    \item {At $r=R$ where $\rho \rightarrow 0$, the viscous acceleration must remain finite (see Eq. \ref{equ:fvis}), hence $q \rightarrow 0$ at the outer boundary. Equation \ref{equ:4-parameter-equations-y4} thus requires}
    \begin{align}
        {2 y_1 + y_3=0.}\label{equ:bc4}
    \end{align}

\end{enumerate}

In the simpler case of $\delta \Omega/\Omega = f_\Omega (\delta r/r)$, there are only 2 equations and 2 boundary conditions. The inner boundary condition is the same as Eq. \eqref{equ:bc1}. The momentum Eq. \eqref{equ:4-parameter-equations-y2} becomes
\begin{align}
    r \frac{\dd y_2}{\dd r} & = \frac{GM\rho}{r p} \left[\left(4+ \frac{r^3 \omega^2}{GM}+ (2f_\Omega  -1) \frac{r^3 \Omega^2}{GM} \right)y_1 \right. \nonumber\\
    &+ \left. \left(1-\frac{r^3 \Omega^2}{GM}\right) y_2 \right].\label{equ:2-parameter-equations-y2}
\end{align}
and hence the outer boundary condition changes to 
    \begin{align}
        \big[4 + \tilde{\omega}^2 + (2 f_\Omega -1) \tilde{\Omega}^2 \big] y_1 + (1 - \tilde{\Omega}^2) y_2 = 0.\label{equ:bc3p}
    \end{align}
In the case of no AM transport with $f_\Omega=-2$, angular momentum conservation is automatically satisfied. 

To summarize, we have two situations with different sets of equations to solve.
\begin{itemize}
    \item The ``full" version including 4 equations (Eqs. \ref{equ:4-parameter-equations-y1}--\ref{equ:4-parameter-equations-y4}) and 4 boundary conditions (Eqs. \ref{equ:bc1}--\ref{equ:bc4}), where perturbed quantities and $\tilde{\omega}$ are all complex. Viscous AM transport is included in this model.
    \item The ``simplified" version including 2 equations (Eqs. \ref{equ:4-parameter-equations-y1}, \ref{equ:2-parameter-equations-y2}) and 2 boundary conditions (Eqs. \ref{equ:bc1}, \ref{equ:bc3p}), where perturbed quantities and $\tilde{\omega}^2$ are all real. Viscous friction is replaced by the prescription that $\delta \Omega/\Omega = f_\Omega (\delta r/r)$.
\end{itemize}

We numerically solved these equations with a shooting method, with details included in Appendix \ref{app:solving-perturbation-equations}. In brief, we first discretize the differential equations at a set of grid points. Then we build a determinant function and ``shoot" (search) over a range of frequency for eigenvalues such that the determinant is equal to zero. Finally, we solve for the corresponding eigenfunctions. The whole process is explained in \cite{TownsendTeitler2013} for the code GYRE. Due to the different equations that we solve, we implemented our own solver. We also introduce a new scheme for solving for complex eigenvalues that is different from \cite{GoldsteinTownsend2020}.

\subsection{High- and Low-viscosity Limits}
\label{sec:perturbations:viscosity}

Physically, in the low viscosity limit ($\tilde{\nu} \to 0$), the viscous AM transport is negligible so each shell of the star conserves angular momentum and requires $\delta(\Omega r^2)=0$, which means that $f_\Omega=-2$ holds everywhere. On the contrary, if $\tilde{\nu} \to \infty$, the viscous force is very significant so perturbations in $\Omega$ become difficult. In this limit, the star is a rigidly rotating body so that $J = I \Omega$, where $I$ is the momentum of inertia. Then AM conservation implies that $\delta \Omega/\Omega = -\delta I/I$, where $\delta I = \int r^2 (2 \delta r/r) dM$. If $\delta r/r$ is an oscillatory function, then $\delta \Omega/\Omega = - \delta I/I \simeq 0.$


\subsubsection{Low-viscosity limit}

As demonstrated above, when $\nu \to 0$, the solutions to the 4-variable equations will reduce to the 2-variable equations with $f_\Omega=-2$. In many stellar models, the viscosity is very low ($\tilde{\nu} \ll 1$), which means $f_\Omega=-2$ can be a good approximation. However, we would still like to calculate mode growth/damping rates in the low viscosity limit, similar to calculations of mode growth/damping rates in the weakly non-adiabatic limit. To compute the imaginary part of eigenvalues of the 4-variable equations, we may study the behavior near the low-viscosity limit.

We start from the $f_{\Omega}=-2$ approximation, where the perturbation equation can be written into a second order differential equation in terms of $y_1$ only:
\begin{align}
    \mathcal{L} y_1 = \lambda w y_1. \label{equ:sturm-liouville}
\end{align}
Here $\mathcal{L}$ is a linear operator which is in the standard Sturm-Liouville form such that $[P(x) u']'+Q(x)u = \lambda w(x) u$, where
\begin{align}
    P(r) & = -\Gamma_1 p r^4;\\
    Q(r) & = -\frac{\dd p}{\dd r} r^3 (3\Gamma_1 -4) - (2f_\Omega+3) \Omega^2 w.
\end{align}
The eigenvalue $\lambda = \omega ^2$ is real, and $w=\rho r^4$ is the weight function. The inner product of two quantities $X$ and $Y$ is defined as
\begin{align}
    \expval{X,Y} = \int_0^{R} X Y \rho r^4 \dd r.
\end{align}

We then write the 4-variable equations (Eqs. \ref{equ:4-parameter-equations-y1}--\ref{equ:4-parameter-equations-y4}) into second order operators and $(y_1, y_3)$ (i.e., by eliminating $y_2$ and $y_4$). The new equations can be written into a succinct matrix form as
\begin{align}
    \begin{pmatrix}
        \mathcal{L} + 2f_\Omega w \Omega^2 & -2 w \Omega^2 \\
        w' \mathcal{L}_1 & w '\mathcal{L}_3
    \end{pmatrix}
    \begin{pmatrix}
        y_1 \\ y_3
    \end{pmatrix}
    = \begin{pmatrix}
        \lambda w & 0 \\
        2\sqrt{\lambda} & \sqrt{\lambda}
    \end{pmatrix}
    \begin{pmatrix}
        y_1 \\ y_3
    \end{pmatrix}. \label{equ:pmatrix}
\end{align}
Here $w' = iq\nu/r^2$, and  $\mathcal{L}_1$ and $\mathcal{L}_3$ are defined as 
\begin{align}
    \mathcal{L}_1 & = -2 r^2 \frac{\dd ^2}{\dd r^2} - 2r \frac{\dd }{\dd r};\\
    \mathcal{L}_3 & = \frac{r^2}{q} \frac{\dd^2}{\dd r^2} + \left(-\frac{r}{q^2} \frac{\dd q}{\dd r} + \frac{1}{q} + 1 \right) r \frac{\dd }{\dd r}.
\end{align}
When $\nu\to 0$, the whole equation will return to the $f_\Omega=-2$ condition.

We perturb the equations away from that solution for a small viscosity by expanding $y_1 \to y_1 + \delta y_1$ and $y_3 \to y_3 + \delta y_3 = f_\Omega y_1 + \delta y_3$, and $\lambda \to \lambda+\delta \lambda = \omega^2 + 2 \omega  \delta \omega $. Above, $\delta y_1$ is composed of a sum of modes with different eigenvalues, i.e., it is orthogonal to $y_1$. Inserting the perturbed quantities into the top row of Eq. \eqref{equ:pmatrix}, keeping the lowest order terms, using $f_\Omega = -2$, and using orthogonality (i.e., $\expval{ y_1, \delta y_1}=0$), we have
\begin{align}
    \delta \omega & = \frac{\expval{y_1, \Omega^2(-2 \delta y_1- \delta y_3)} }{ \omega \expval{y_1, y_1}}.\label{equ:delta_omega}
\end{align}

In the limit $\nu \to 0$,
the perturbed eigenvectors should be linearly proportional to $\nu$, so we assume that
\begin{align}
    2 \delta y_1 + \delta y_3 = \alpha_\nu \nu,\label{equ:alpha_nu}
\end{align}
where $\alpha_\nu$ is a coefficient of order unity. Then in the $y_3$-equation (bottom row of Eq. \ref{equ:pmatrix}), we keep only terms with one power of $\nu$, which yields
\begin{align}
    \frac{i q \nu}{r^2}(\mathcal{L}_1 -2\mathcal{L}_3)y_1 = \omega (2 \delta y_1 + \delta y_3).\label{equ:y_3-equation-lowest-order}
\end{align}
By combining Eqs. \eqref{equ:delta_omega} and \eqref{equ:y_3-equation-lowest-order}, we find
\begin{align}
    \delta \omega =  -i\frac{\expval{y_1, (q \Omega^2 \nu /r^2) (\mathcal{L}_1 -2\mathcal{L}_3)y_1} }{ \omega^2 \expval{y_1, y_1}}.\label{equ:low-viscosity-perturbation}
\end{align}
Note that the perturbation to the mode frequency is purely imaginary, such that the real part of $\omega$ is unchanged, and Eq. \eqref{equ:low-viscosity-perturbation} is the mode growth rate in the low viscosity limit. This is analogous to a work integral in the weakly non-adiabatic limit.

The above result gives us a new method to estimate mode growth rates without solving the whole complex 4-variable equation. Since $\nu$ can be very small (e.g., $\tilde{\nu}\sim 10^{-5}$) and is in the denominator of Eq. \eqref{equ:4-parameter-equations-y4}, numerical problems can arise in the full set of equations. On the contrary, solving $y_1$ using $f_\Omega=-2$, and then computing a mode growth rate from Eq. \eqref{equ:low-viscosity-perturbation} is much easier and usually more accurate. We numerically validate this method in the following section.

We may also relate $\delta \omega$ to the work done by the viscous force due to small perturbations. Since $\mathcal{L}_1 y_1+ \mathcal{L}_3y_3 = r \dd y_4/ \dd r=r \delta f_\phi /(\Omega q \nu)$, and $\delta v_\phi = r\delta \Omega + \delta r \Omega = -\Omega r y_1$, we find
\begin{align}
    \delta \omega = \frac{i}{4\pi \omega^2 \expval{y_1, y_1}} \int_0^ R \delta f_\phi \delta v_\phi \dd M = \frac{i}{4\pi \omega^2 \expval{y_1, y_1}} \frac{\dd W}{\dd t} \, ,
\end{align}
where $dW/dt$ is the work done on the star by the viscous force. If $\dd W/\dd t>0$, the viscous force does work on the star, causing the mode to grow such that $\Im \delta \omega>0.$

It is useful to evaluate Eq. \eqref{equ:low-viscosity-perturbation} in certain limits. Note that
\begin{align*}
        \mathcal{L}_1 -2\mathcal{L}_3  = -2 \left[r \frac{\dd }{\dd r}\left(\frac{r}{q}\frac{\dd }{\dd r} \right) + \frac{\dd }{\dd r} \left(r^2 \frac{\dd }{\dd r}\right)  \right].
\end{align*}
We plug this into Eq. \eqref{equ:low-viscosity-perturbation} and integrate by parts. If we adopt the WKB approximation such that only derivatives of $y_1$ are kept, we find
\begin{align}
    \Im \delta {\omega} \propto \int_0^R \rho r^2 \Omega^2 \nu (1+q) \left| r \frac{\dd y_1}{\dd r} \right|^2 \dd r. \label{equ:low-viscosity-perturbation-simplified}
\end{align}
Then in the $q\to 0$ limit,
\begin{align}
    \Im \delta \omega \propto \frac{-2}{\omega^2} \frac{\expval{\nu \Omega^2, |\dd y_1 /\dd r|^2}}{\expval{y_1, y_1}}<0.
\end{align}
which means the oscillation modes are always stable in the low-shear and WKB limits, and they cannot grow in the absence of shear. This makes sense, as shear is the energy source for viscously driven modes. 

We see from Eq. \eqref{equ:low-viscosity-perturbation-simplified} that  a necessary but not sufficient condition for an unstable mode in the WKB limit is $q<-1$ somewhere within the star. High-order modes only grow if the shear is sufficiently negative, i.e., with large degrees of outwardly decreasing differential rotation. This type of shear often arises in stellar models where contracting inner layers spin up and composition gradients impede angular momentum transport, producing large negative shears.

\section{Polytropic Models}
\label{sec:polytropic}

To understand the possibility of viscous mode excitation in stars, we first examine simple polytropic stellar models. These models are based on a polytropic equation of state, i.e., $p = K \rho^{\gamma}$, where $\gamma$ is a constant polytropic index and $K$ is a constant. Hydrostatic equilibrium is modified by the centrifugal force, and we still assume the rotational rate is well below the {critical} limit so latitudinal dependence of density and pressure is ignored. To solve for the stellar structure, we first introduce new variables $\vartheta$ and $\xi$, defined as $\rho \propto \vartheta^{1/(\gamma -1)}$, $r = r_\gamma \xi$, and $r_\gamma^2 = \gamma p_{\rm c}/[4\pi (\gamma -1) G \rho_{\rm c}^2]$, where $p_{\rm c}$ and $\rho_{\rm c}$ are pressure and density at the center.
Hydrostatic equilibrium then requires
\begin{align}
    \frac{1}{\xi^2} \frac{\dd }{\dd \xi} \left(-\xi^2 \frac{\dd \vartheta}{\dd \xi} + \Lambda \xi^3 \right)  = \vartheta^{1/(\gamma-1)}.\label{equ:lene-emden_rotation}
\end{align}
Here $\Lambda = \Omega^2/(4\pi G \rho_{\rm c})$ is the additional term that modifies the Lane-Emden equation \cite[]{Lane1870} to include the centrifugal force. Due to the additional term, the radius of the star ($\xi_1$, the first positive root to $\vartheta(\xi)=0$) will change.

\subsection{Low-viscosity Limit}
\label{sec:polytropic:low-vis}

\begin{figure}
    \centering
    \includegraphics[width=\linewidth]{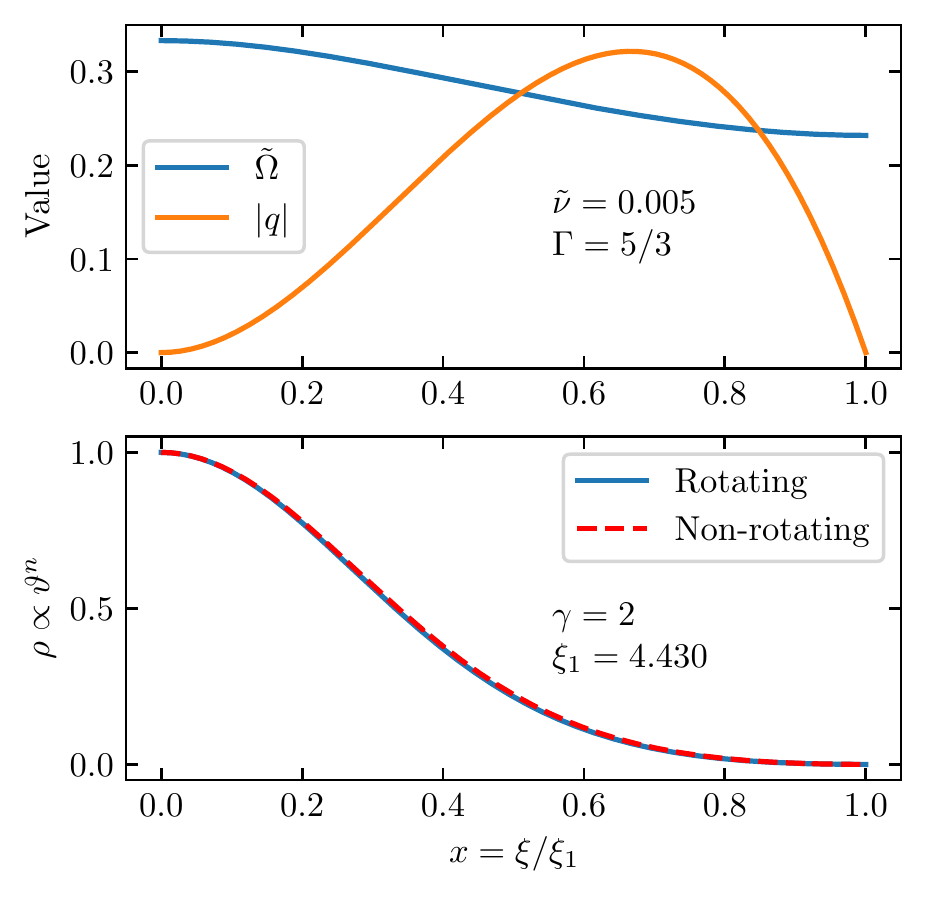}
    \caption{A rotating stellar model with polytropic index $\gamma=3/2$, adiabatic index $\Gamma=5/3$, and a constant small viscosity $\tilde{\nu}=0.005$.
    \textbf{Top}: rotational profile and shear of the star used for calculations in Fig. \ref{fig:polytrope-modes-low-viscosity}, plotted as a function of normalized radial coordinate.
    \textbf{Bottom}: density profile of the star when it is non-rotating (red dashed line) or rotating as shown in the top panel (blue line), highlighting the change caused by centrifugal support.
    }
    \label{fig:polytrope-low-viscosity}
\end{figure}

\begin{figure}
    \centering
    \includegraphics[width=\linewidth]{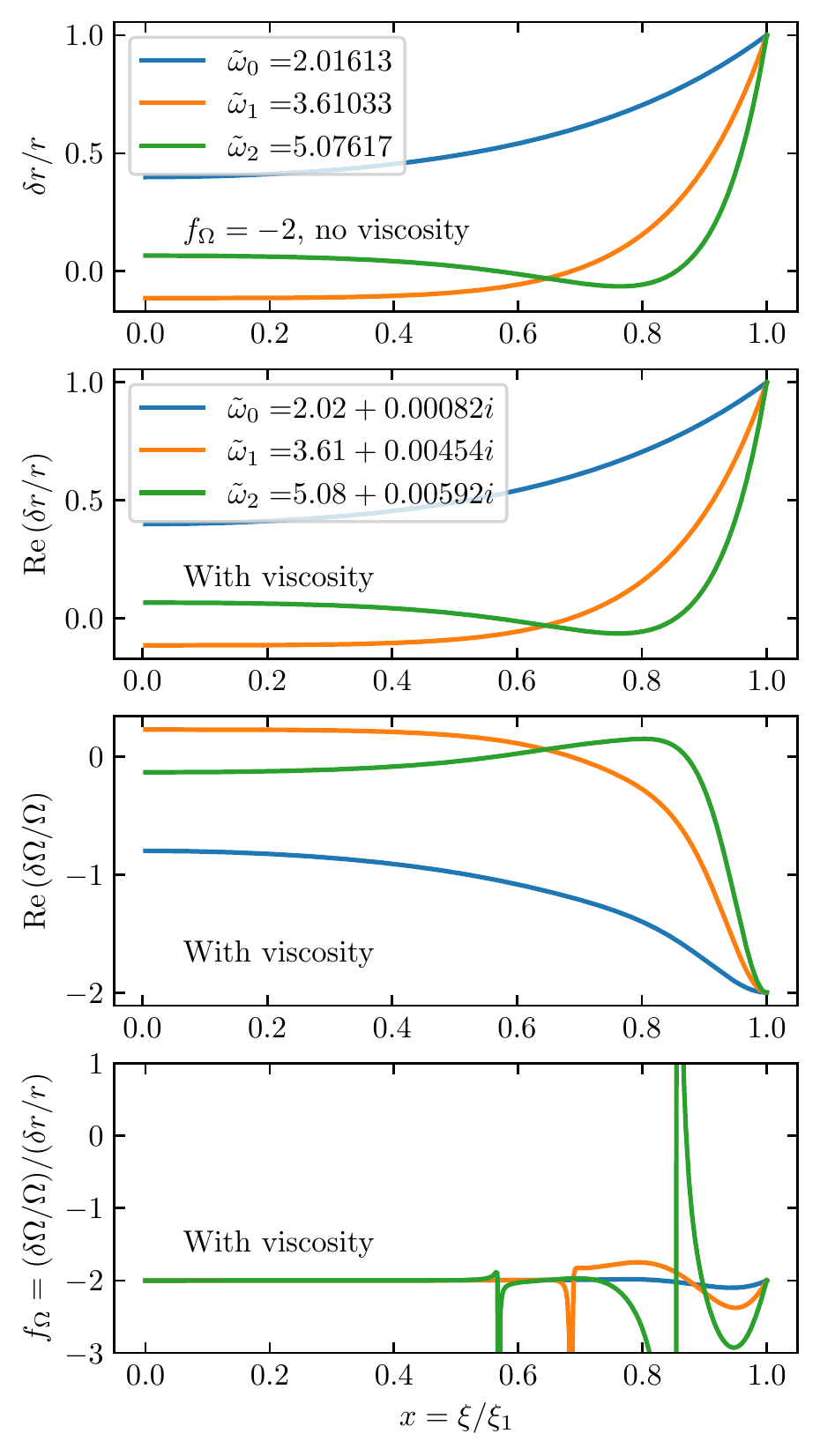}
    \caption{Radial oscillation modes of the low viscosity polytropic model shown in Fig. \ref{fig:polytrope-low-viscosity}. \textbf{Top panel}: radial displacement eigenfunctions solved by fixing $f_\Omega \equiv (\delta \Omega/\Omega)/(\delta r/r)=-2$ (i.e., no viscosity), with associated frequencies shown by the legend.
    \textbf{Middle two panels}: 
    radial displacement and rotation perturbation eigenfunctions for modes including a small viscosity of $\bar{\nu}=0.005$. The mode frequencies are now complex, with positive imaginary components corresponding to excited modes. The real parts of the mode frequencies and eigenfunctions are very close to the $f_\Omega=-2$ approximation.
    \textbf{Bottom panel}: ratios between $\delta \Omega/\Omega$ and $\delta r/r$ for the three modes shown in the middle two panels. We see that $f_\Omega=-2$ is a good approximation, except near nodes of the eigenfunctions.
    }
    \label{fig:polytrope-modes-low-viscosity}
\end{figure}

\begin{table}
    \centering
    \begin{tabular}{llllll}
    \toprule
    $n$ & $\tilde{\omega}$ (no vis.)&  $\Re \tilde{\omega}$ (vis.) & $ \Re (\Delta \tilde{\omega})$ & $\Im \tilde{\omega}$ (vis.) & $\Im \delta \tilde{\omega}$\\
    \midrule
0 & 2.016127 & 2.016074 & 0.000052 & 0.000824 & 0.000688\\
1 & 3.610328 & 3.610890 & 0.000562 & 0.004543 & 0.003592\\
2 & 5.076167 & 5.079960 & 0.003793 & 0.005925 & 0.005396\\
    \bottomrule
    \end{tabular}
    \caption{Mode frequencies of the polytropic star as shown in Fig. \ref{fig:polytrope-low-viscosity}. Columns in this table are: 1) mode radial order; 2) mode frequency calculated with $f_\Omega=-2$ approximation; 3) real part of mode frequency including low viscosity; 4) relative difference between frequencies with and without viscosity; 5) imaginary part of mode frequency including low viscosity; 6) imaginary part of mode frequency calculated from the low-viscosity limit (Eq. \ref{equ:low-viscosity-perturbation}). 
    }
    \label{tab:eigenvalues}
\end{table}

We first focus on a $\gamma=3/2$ polytropic model with a constant low viscosity, $\tilde{\nu}=0.005$. We also set a constant adiabatic index, $\Gamma=5/3$. {The rotational profile is chosen such that derivatives of $\Omega$ are zero at both the core and surface:
\begin{align}
    \Lambda = A\exp[-f_A\left(\frac{\xi_1\xi^2}{2}-\frac{\xi^3}{3}\right)].\label{equ:polytrope-rotation-profile}
\end{align}
We choose $A\approx 0.0032$, $f_A=0.05$, and $\xi_1 \approx 4.429861$, which is self-consistently the radius of the polytropic star. As a result, we have $q =(f_A/2)\cdot \xi^2 (\xi-\xi_1)$, which satisfies $q\propto \xi^2$ near the center and $q=0$ at the surface.} The density profile of the distorted polytrope is shown in Fig. \ref{fig:polytrope-low-viscosity}.

We then calculate the oscillation modes using the full set of equations (Eqs. \ref{equ:4-parameter-equations-y1}--\ref{equ:4-parameter-equations-y4}) and compare them with the $f_\Omega=-2$ approximation (Eq. \ref{equ:2-parameter-equations-y2}). Modes are distinguished with a radial order $n$, which is the number of nodes (or roots) of $\delta r/r$, and modes with higher $n$ have higher frequency. In Fig. \ref{fig:polytrope-modes-low-viscosity}, we can see that the $\delta r/r$ eigenfunctions are quite similar under the two conditions. We also compare the ratio between perturbations in rotational rate and radius, which show reasonable agreement with $f_\Omega=-2$ in much of the star for all three modes. Deviation from $f_\Omega = (\delta \Omega/\Omega)/(\delta r/r) \approx -2$ occurs near nodes where $\delta r/r \rightarrow 0$.

Table \ref{tab:eigenvalues} shows the mode frequencies using $f_\Omega = -2$ and those computed for low viscosity. {In this case, all three modes have positive imaginary components of their eigenfrequencies, meaning all three modes are overstable.}. This proves that viscosity coupled with differential rotation can (in principle) excite stellar oscillation modes. For each mode, the real part of the eigenvalue is almost the same as the $f_\Omega=-2$ approximation, and the absolute difference ($\Re (\Delta \tilde{\omega})$) is at least three times smaller than the imaginary part ($\Im \tilde{\omega}$), which is expected from the linear perturbation in viscosity developed in Sec. \ref{sec:perturbations:viscosity}. We also compare $\Im \tilde{\omega}$ with the growth rate in the low-viscosity limit (Eq. \ref{equ:low-viscosity-perturbation}) in Tab. \ref{tab:eigenvalues}. For all three modes, we see $\Im \tilde{\omega} \approx \delta \tilde{\omega}$ holds, and the relative difference is $\lesssim 20\%$. This demonstrates the value of the perturbation analysis of Section \ref{sec:perturbations:viscosity} such that we may calculate mode growth/damping rates using Eq. \eqref{equ:low-viscosity-perturbation} without solving the full set of complex equations (Eqs. \ref{equ:4-parameter-equations-y1}--\ref{equ:4-parameter-equations-y4}). 


\subsection{High-viscosity Limit}
\label{sec:polytropic:high-vis}

\begin{figure}
    \centering
    \includegraphics[width=\linewidth]{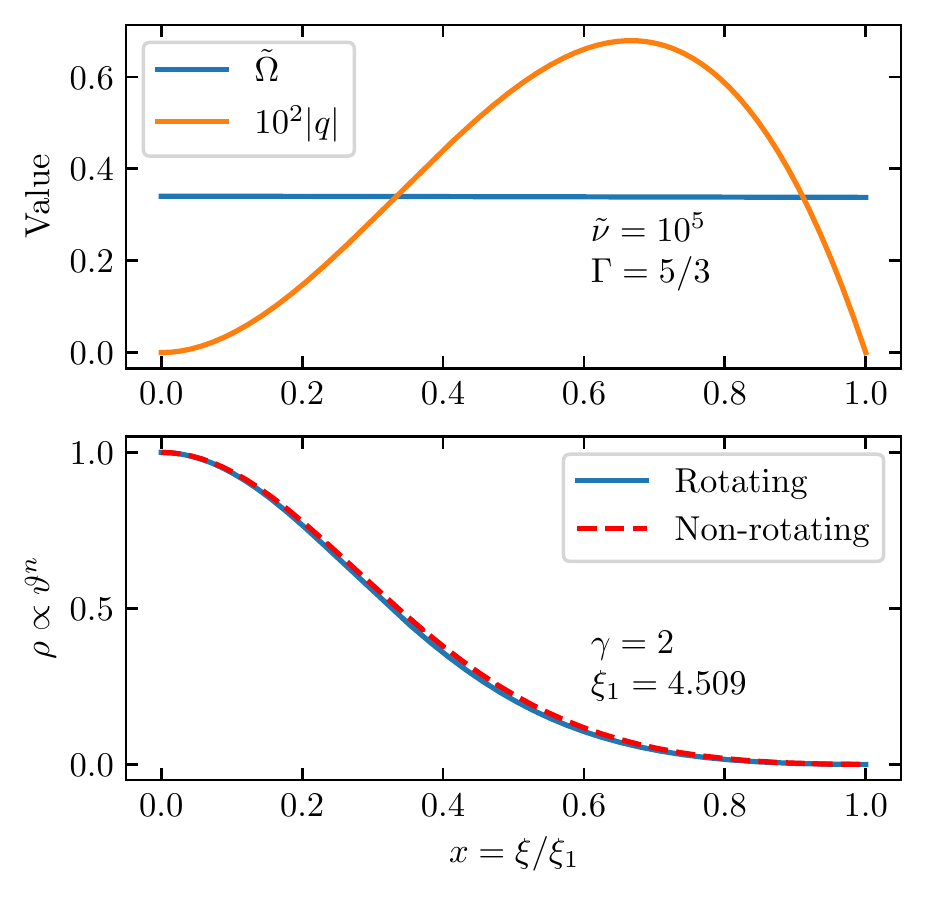}
    \caption{
    The same as Fig. \ref{fig:polytrope-low-viscosity}, but this time for a polytropic model with high viscosity $\tilde{\nu}=10^{5}$ and nearly flat rotation profile.
    }
    \label{fig:polytrope-high-viscosity}
\end{figure}

\begin{figure}
    \centering
    \includegraphics[width=\linewidth]{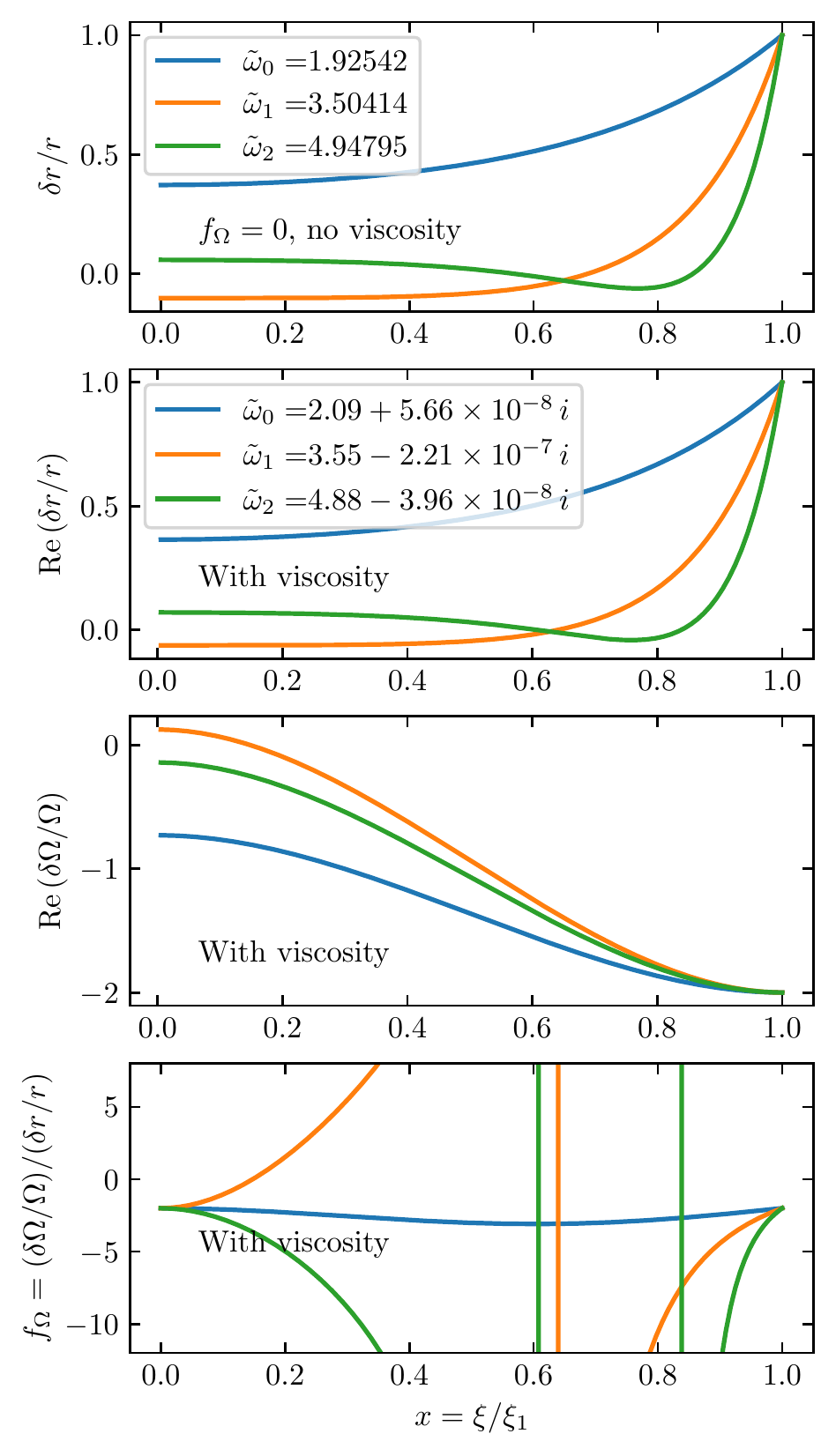}
    \caption{Similar to Fig. \ref{fig:polytrope-modes-low-viscosity}, but for the high-viscosity model shown in Fig. \ref{fig:polytrope-high-viscosity}. Due to the high viscosity, we first solve for oscillation modes with $f_\Omega=0$ (top panel), and we then solve the full set of equations including viscous forces (middle two panels). The real parts of eigenvalues and eigenfunctions in the two cases are very similar, and we find both damped and growing modes. Ratios between the rotation and radius perturbations are near $f_\Omega=0$ in the outer part of the star for higher order modes (bottom panel).
    }
    \label{fig:polytrope-modes-high-viscosity}
\end{figure}

It is also instructive to examine the high-viscosity limit. Fig. \ref{fig:polytrope-high-viscosity} shows a $\gamma=3/2$ polytropic model with a constant $\Gamma=5/3$. {The rotation profile follows Eq. \eqref{equ:polytrope-rotation-profile} with the same $A$, but with $f_A=10^{-3}$ and  $\xi_1=4.509417$ such that $q$ approaches zero at each boundary.} As a result, there is a nearly constant rotational rate $\tilde{\Omega}\sim 0.35$ so that $q \approx 0$. The rotation causes the radius $\xi_1$ to slightly increase, along with a slight deviation to the density profile. We assume the star has constant and very high viscosity of $\tilde{\nu}=10^{5}$, such that $|q\tilde{\nu}|$ is very large.

Since we are considering a case with $q \ll 1$, and $q \nu \gg 1$ solutions require a special form of $\delta \Omega/\Omega$ such that the last term of equation 
\ref{equ:euler-horizontal} remains finite. {This requires $\delta \Omega/\Omega \approx a + \int (b q/r)dr \simeq a + b [\Omega(r) - \Omega(0)]/\Omega(0)$, where $a$ and $b$ are constants. The boundary values require $a \simeq -2 y_1(0)$ and $b \simeq -\Omega(0) [2 + a]/[\Omega(r)-\Omega(0)]$. In this limit, the magnitude of $q$ does not affect the solution (though its functional form does), and the growth rate depends on the magnitude of ${\bar \nu}$ but not the magnitude of $q$.} 

Additionally, conservation of total angular momentum requires $\int dM r^2 \delta \Omega = - \int dM 2 \Omega r^2 (\delta r/r)$. For high-order modes with an oscillatory $\delta r/r$, the right hand side is small and we expect the average value of $\delta \Omega/\Omega$ to be small, even though the local value of $\delta \Omega/\Omega$ is not small. {Hence, using $f_\Omega = 0$ may be a good approximation even though the actual values of $f_\Omega$ are not typically close to zero.}


We calculate the radial oscillation modes with high viscosity from Eqs. (\ref{equ:4-parameter-equations-y1}--\ref{equ:4-parameter-equations-y4}), in addition to calculating them with the $f_\Omega=0$ approximation. In Fig. \ref{fig:polytrope-modes-high-viscosity}, we compare oscillation modes solved under the two conditions. We see that the eigenvlaues are quite close, and the shapes of $\delta r/r$ eigenfunctions are also very similar. The agreement is worst for the $n=0$ mode, as expected because it produces a significant change in the moment of inertia so we do not expect $f_\Omega = 0$ to be a very good approximation. 
{In this case, only one of the modes computed (the $n=0$ mode) is found to be overstable, while the $n=1$ and $n=2$ modes are damped. 
As expected, we find that $\delta \Omega/\Omega$ has the same shape for each mode (regardless of radial order), as explained above. We see that $f_\Omega$ is not generally close to zero, but for high-order modes it oscillates around zero, such that the $f_\Omega=0$ approximation may be appropriate.}

\section{Realistic Models}
\label{sec:mesa}

\begin{figure}
    \centering
    \includegraphics[width=\linewidth]{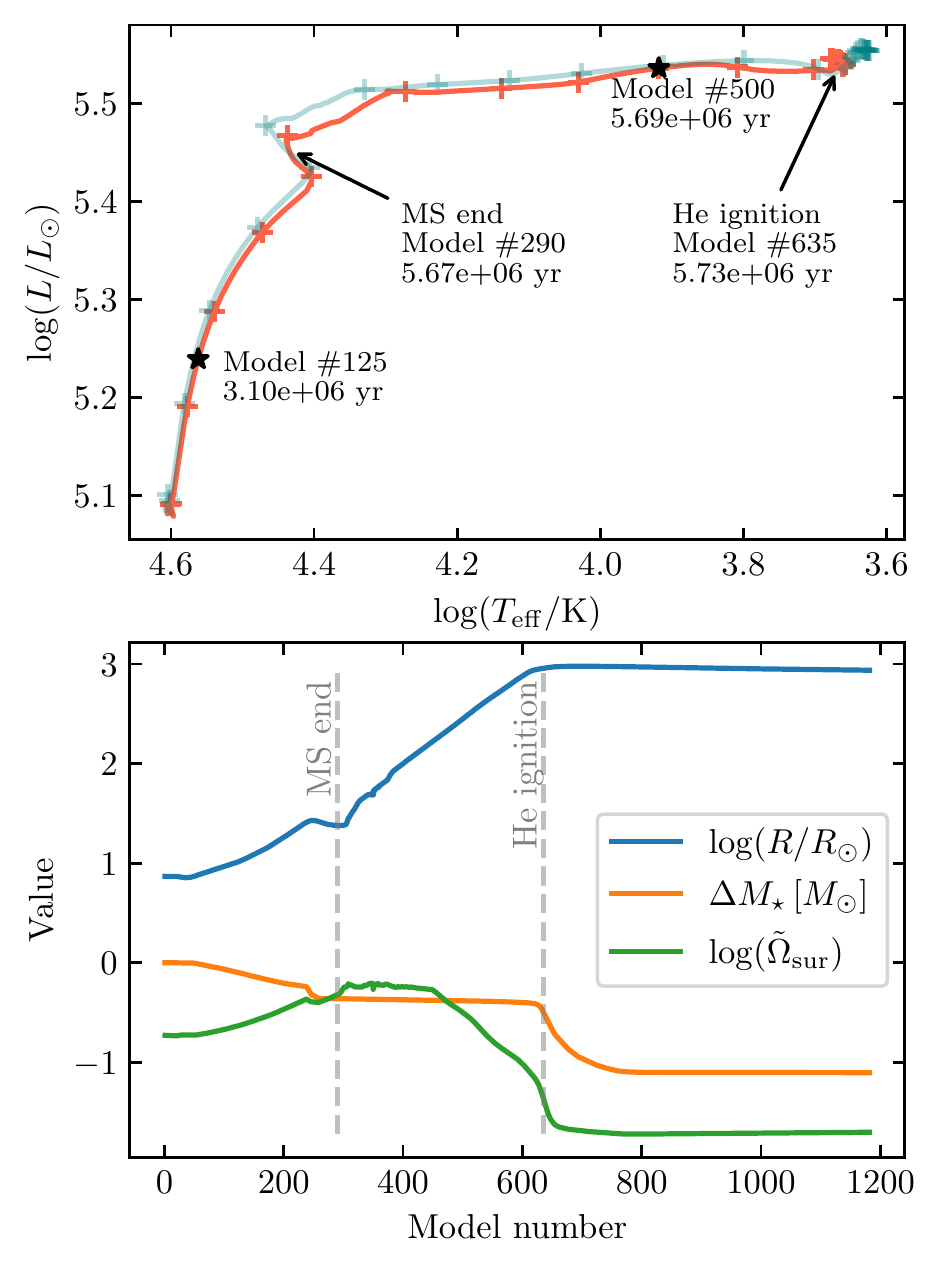}
    \caption{Evolution of a $30 \, M_\odot$ MESA model.
    \textbf{Top}: Evolution through the HR diagram, with labels marking the end of the main sequence and the start of helium burning. Every 50th model is marked with a cross. For comparison, we also show the evolution track of a non-rotating but otherwise identical model (green curve).
    \textbf{Bottom}: Evolution of surface radius, mass lost via winds, and surface rotational rate of the star. Note that the surface rotation rate approaches {the critical rate} ($\tilde{\Omega}_{\rm sur} \! \sim \! 1$) just after the end of the main sequence.
    }
    \label{fig:mesa-model}
\end{figure}

In this section, we perform our calculations for more realistic massive star models generated using the MESA stellar evolution code \citep{paxton:11,paxton:13,paxton:15,paxton:18,PaxtonSmolec2019}.
Our chosen model has an initial mass of $30\,M_\odot$ at solar metallicity and the initial rotational profile is flat with $\tilde{\Omega}=0.25$. Rather than using a realistic AM viscosity prescription like \citet{FullerPiro2019}, we choose a simplified viscosity prescription of $\tilde{\nu} = 10^{-5} (r/R)^2$, which is quite low even at the surface. We have found that realistic viscosity prescriptions generate abrupt changes in $\Omega$ with radius which are numerically problematic. Our choice of viscosity enforces nearly rigid rotation during the main sequence but allows for large amounts of differential rotation in the post-main sequence, as often occurs in stellar models with more realistic AM transport prescriptions.

\begin{figure}
    \centering
    \includegraphics[width=\linewidth]{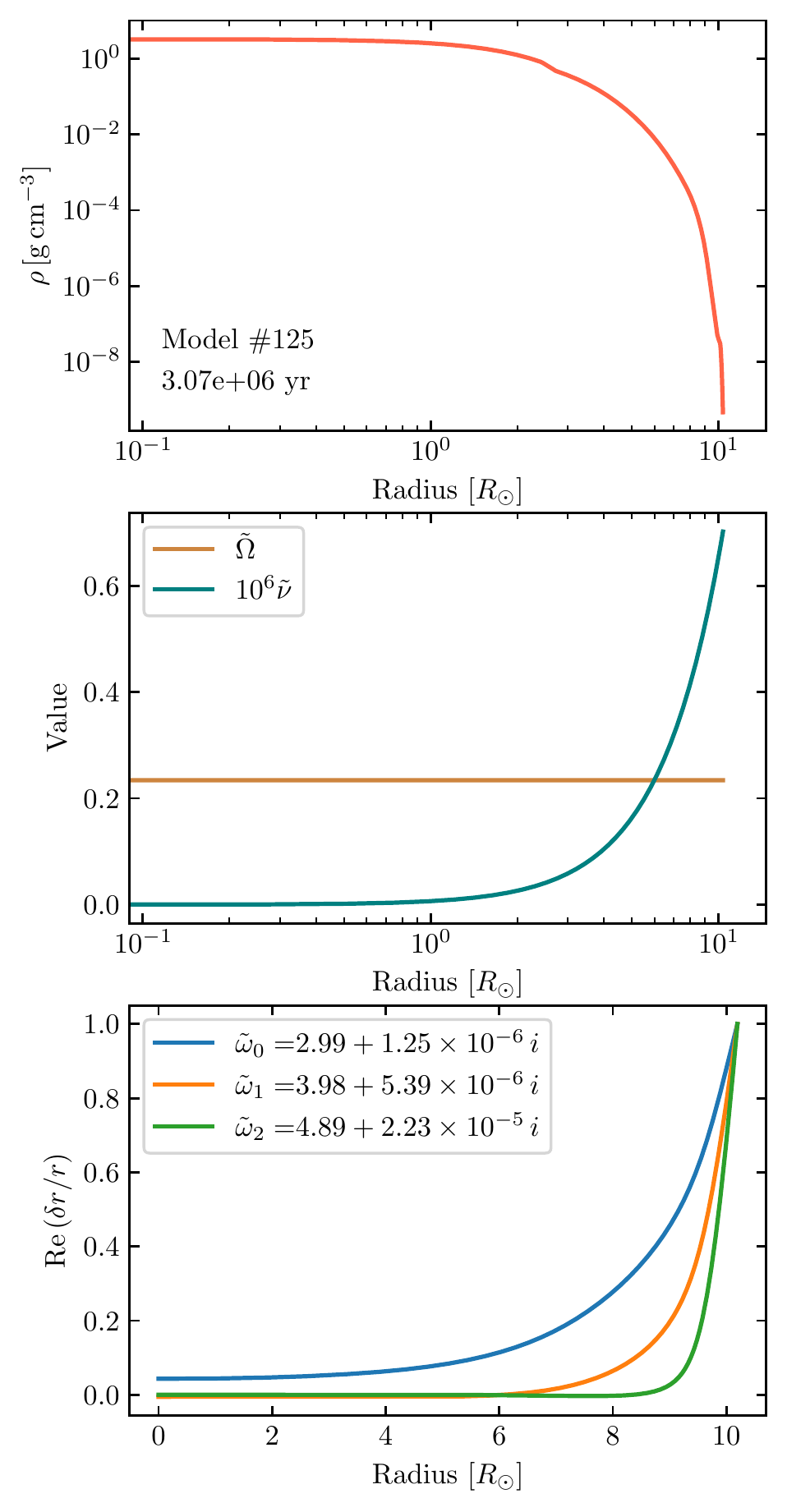}
    \caption{{\bf Top two panels:} Density, rotation, and viscosity profiles of our $30 \, M_\odot$ model on the main sequence (Model \#125). {\bf Bottom panel:} Eigenfunctions of the three lowest order radial modes (with the legend indicating their frequencies), which are destabilized by viscous driving.}
    \label{fig:model-before-main-sequence}
\end{figure}

\begin{figure}
    \centering
    \includegraphics[width=\linewidth]{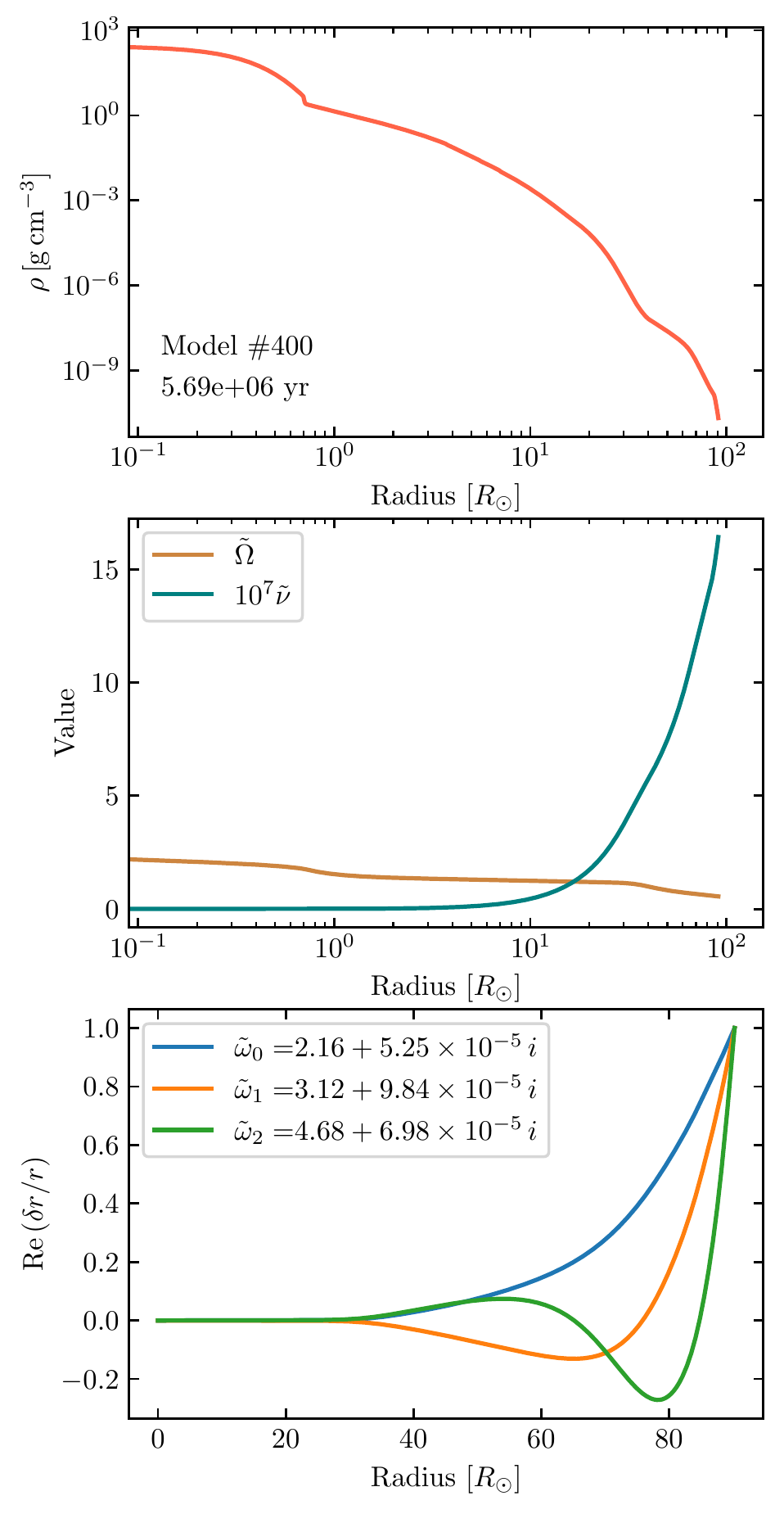}
    \caption{Similar to Fig. \ref{fig:model-before-main-sequence}, but for our $30 \, M_\odot$ model while it is crossing the Hertzsprung gap (Model \#400).  The three lowest-order radial oscillations are destabilized by viscous driving.}
    \label{fig:model-after-main-sequence}
\end{figure}

The evolution track of the star is shown in Fig. \ref{fig:mesa-model}. At $t=5.67\times 10^6\,{\rm yr}$, the star leaves the main sequence when the central hydrogen mass fraction is below 0.01\%.
After the main sequence phase, the star expands its radius by nearly 2 orders of magnitude, and the surface rotation rate drops. However, the dimensionless rotation rate $\tilde{\Omega}$ initially increases to a value near unity (see explanation in \citealt{ZhaoFuller2020}) as the star crosses the Hertzsprung gap, and then decreases to small values when the star becomes a convective red supergiant. Near $t=5.73\times10^{6}\,{\rm yr}$ the star begins to burn helium in the core (the central helium mass fraction is below 90\%). 
Figs. \ref{fig:model-before-main-sequence} and \ref{fig:model-after-main-sequence} show the stellar structure (density, rotation rate, and viscosity profile) for a main sequence model and an HR gap model, respectively. The model crossing the HR gap has a larger radius and its mass is more concentrated at the center of the star (within $\sim \! 1 R_\odot$, while the outer radius is $\sim \! 200\,R_\odot $). The HR gap model also features a large degree of differential rotation, whereas the main sequence model is nearly rigidly rotating.
For comparison we also show the evolution track of a star with the same initial setups but without rotation in Fig. \ref{fig:mesa-model}.

The bottom panels of Figs. \ref{fig:model-before-main-sequence} and \ref{fig:model-after-main-sequence} show the eigenfunctions and growth rates of the first three radial modes in the main sequence and HR gap models. The eigenfunctions are more localized to the outer layers of the star than the polytropic models in Section \ref{sec:polytropic}, as expected due to the large relative density contrast in the realistic stellar models. Most importantly, we see that the growth rates of these modes can be negative or positive (more often the latter), indicating that radial modes can in principal be vsicously excited in massive stars.

\begin{figure}
    \centering
    \includegraphics[width=\linewidth]{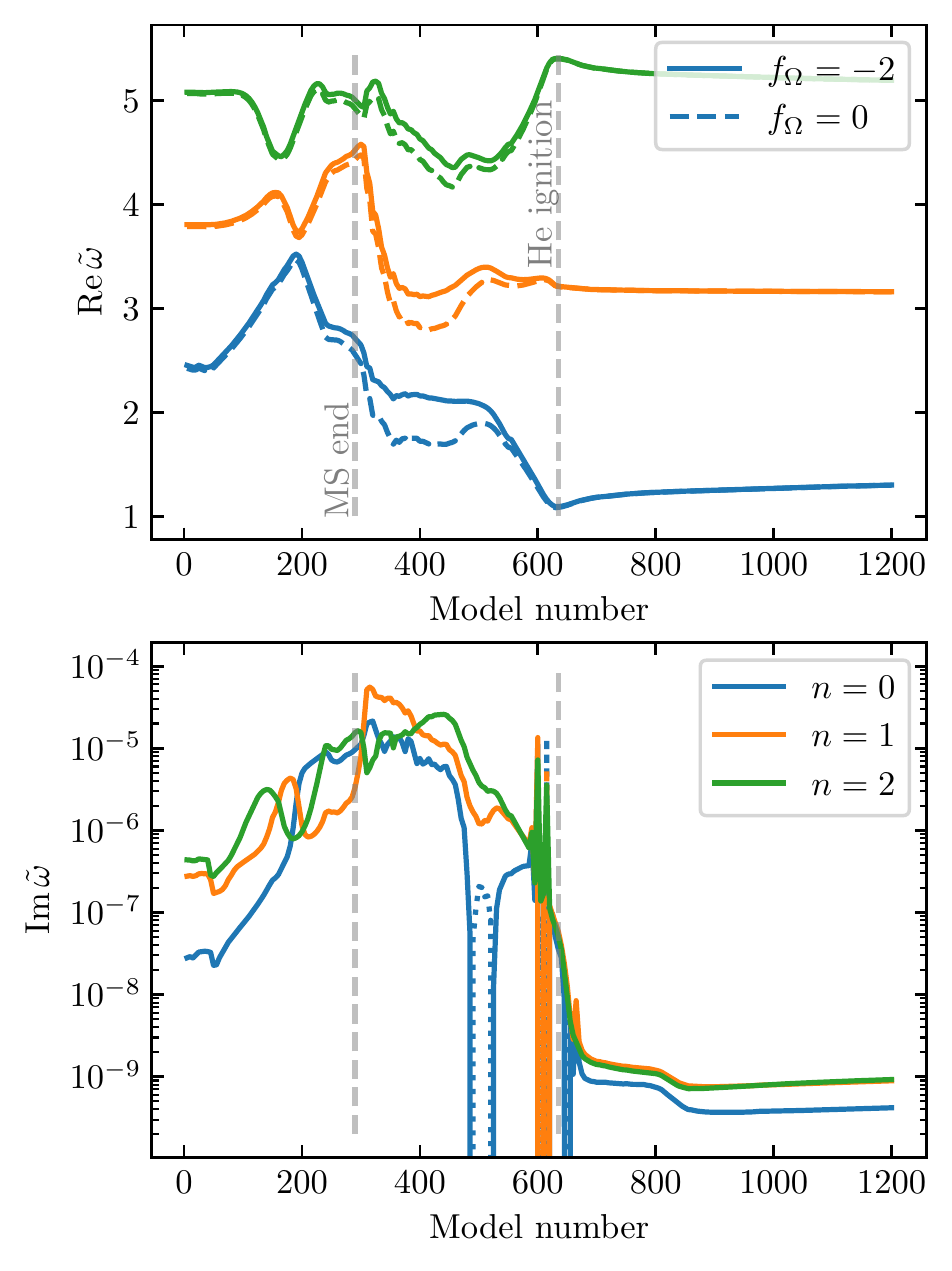}
    \caption{Evolution of low-order radial mode frequencies of our $30 \, M_\odot$ model. 
    \textbf{Top}: Mode frequencies computed with the low-viscosity approximation ($f_\Omega=-2$, solid lines) or high-viscosity approximation ($f_\Omega=0$, dashed lines).
    \textbf{Bottom}: Imaginary parts of the mode frequencies, computed in the low-viscosity limit. During the evolution the imaginary parts are mostly positive (solid lines) and the oscillation modes are unstable, though in rare cases they are negative (dotted lines).
    }
    \label{fig:mesa-eigenvalues}
\end{figure}

\begin{figure}
    \centering
    \includegraphics[width=\linewidth]{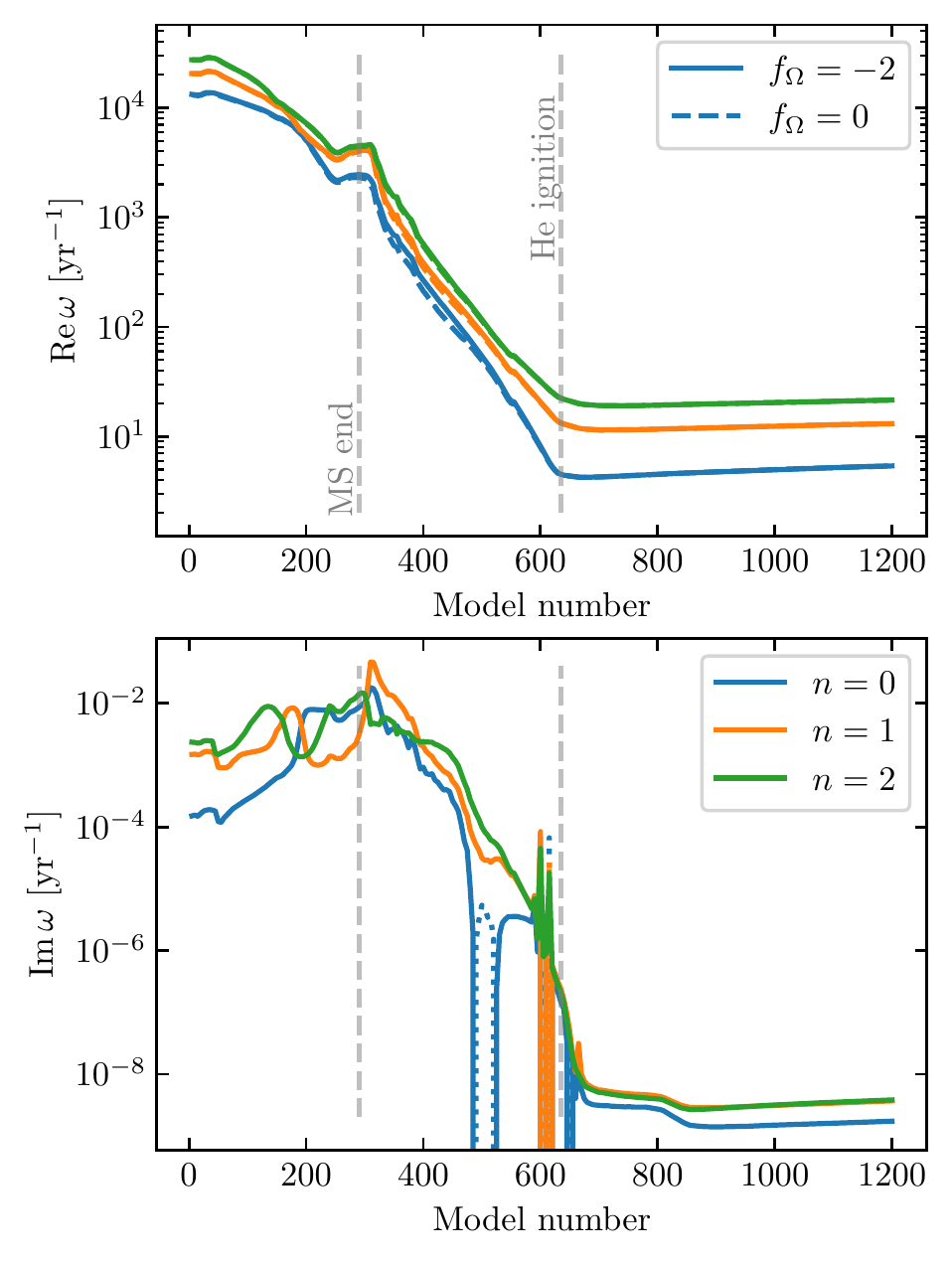}
    \caption{Same as Fig. \ref{fig:mesa-eigenvalues}, but mode frequencies are in physical units (${\rm yr}^{-1}$). The mode growth rate is typically $\sim \! 10^{-4}$ -- $10^{-2}\,{\rm yr}^{-1}$ during the main sequence and blue side of the Hertzsprung gap, but it quickly drops as the star expands into a red supergiant. 
    }
    \label{fig:mesa-eigenvalues-physical}
\end{figure}

Next, we calculate the viscous growth/damping rates as a function of evolutionary state of our model. Due to the low viscosity, we use the $f_\Omega=-2$ approximation for the real part of the eigenvalues, and calculate the imaginary part based on Eq. \eqref{equ:low-viscosity-perturbation} (see Sec. \ref{sec:perturbations:viscosity}). We also present eigenvalues with $f_\Omega=0$ for comparison. Fig. \ref{fig:mesa-eigenvalues} and \ref{fig:mesa-eigenvalues-physical} show the evolution of eigenvalues of the three lowest order radial oscillation modes as a function of evolutionary state. The real parts of the eigenvalues are always larger than $\tilde{\omega} > 1$, even for efficient AM tranpsort from the $f_\Omega=0$ limit. This means that the stars are never dynamically unstable to radial perturbations as suggested by \cite{ZhaoFuller2020}, even with efficient AM transport during the pulsation cycle. 

Instead, we find that modes are usually overstable due to viscous driving, with small (but positive) imaginary components of their frequencies.
Just after the end of the main sequence, the imaginary part of $\tilde{\omega}$ peaks at $\sim \! 10^{-5}$. This corresponds to physical growth times of $\sim$100 years. This makes sense, as the level of differential rotation is smaller on the main sequence (producing low growth rates), and the surface rotation rate is smaller during helium burning (again producing low growth rates). The growth times of $\sim \! 100$ years in the HR gap are shorter than the star's evolutionary time at this phase, potentially allowing viscously driven modes to grow to large amplitudes.


\section{Discussion}
\label{sec:discussion}

\subsection{Observational Implications}

We have demonstrated the possibility of a new source of instability of massive stars, that of viscous overstabilities in differentially rotating stars. The instability utilizes an effective AM viscosity to tap the energy stored in differential rotation to drive acoustic oscillation modes. Though our study focused on low-order radial oscillation modes, it is possible that other types of modes (e.g., high-order radial acoustic modes, non-radial acoustic modes, gravity modes, etc.) could be driven by the viscous instability as well. At the moment it is unclear which types of modes would grow fastest or be most likely to be observed. A comprehensive investigation should include non-adiabatic effects in the perturbation analysis to determine whether viscous driving can dominate over non-adiabatic sources of driving/damping. 

In none of our models did we find a direct centrifugal instability as proposed by \citealt{ZhaoFuller2020} based on a local analysis. In other words, we never found instabilities where {$\Re \omega^2 < 0$}. Instead, we showed that viscous overstability, with {$\Im\omega>0$}, is prevalent in our massive stellar models. If viscous instability occurs in Nature, it would then be most likely to be observed as a stellar pulsation, similar to thermally or convectively driven acoustic/gravity modes. Unlike direct centrifugal instability, it seems less likely that viscous overstability would result in global outbursts and/or masss loss from massive stars like those observed in LBVs. 

Our study focused on massive stellar models, but viscous instabilities could occur in any type of differentially rotating star. For example, low-mass red giant stars are known to be strongly differentially rotating through asteroseismology (e.g., \citealt{beck:12}), and could potentially harbor viscously driven modes.

\subsection{Viscosity}

In this work, we have included a viscous AM transport term, showing that this form of AM transport can destabilize stellar oscillations. However, we have not specified the actual source of viscosity. The microscopic viscosity will almost always be too small to produce any achievable driving. An effective convective viscosity might be the best candidate, naively producing a viscosity of order $\nu_{\rm con} \sim v_{\rm con} \ell_{\rm con}$, where $v_{\rm con}$ and $\ell_{\rm con}$ are the typical convective velocity and mixing length. Magnetic torques can also behave similarly to an effective viscosity of strength $v_{\rm AM} \sim B_r B_\phi/(\rho q \Omega)$ (e.g., \citealt{spruit:02}). Even sluggish convection or moderate magnetic fields can produce torques orders of magnitude larger than those due to microscopic viscosity, so these possibilities should be investigated in more detail. 

Additionally, our calculations have not included a Lagrangian perturbation to the viscosity, $\delta \nu$, since it is not clear how to calculate this term. Future calculations focused on specific sources of AM viscosity should be used to estimate $\delta \nu$, and to determine whether this term can affect mode growth rates.

\subsection{Non-adiabatic, Non-linear, and Non-radial Modes}

Since ours is one of the first investigations into viscously driven instabilities, we have neglected non-adiabatic effects for simplicity. We have focused on low-order acoustic modes where non-adiabatic effects are minimized, but future work should investigate these effects, which could dominate the driving and/or damping of acoustic modes. For instance, it is well known that massive, very luminous stars could undergo strange mode instabilities \citep{glatzel:94,papaloizou:97,saio:97}. These pulsations modes are partially trapped in the surface layers of the star and have very short growth times. In contrast, the low-order acoustic modes examined here extend deeper into the star and have longer growth/damping times. The non-linear evolution of strange mode instability has been investigated by \citep{yadav:17}, who finds it can cause large expansion of the envelope and a decrease of effective temperature, qualitatively similar to S Doradus outbursts. The non-linear evolution of viscously driven modes is very unclear.

\begin{figure}
    \centering
    \includegraphics[width=\linewidth]{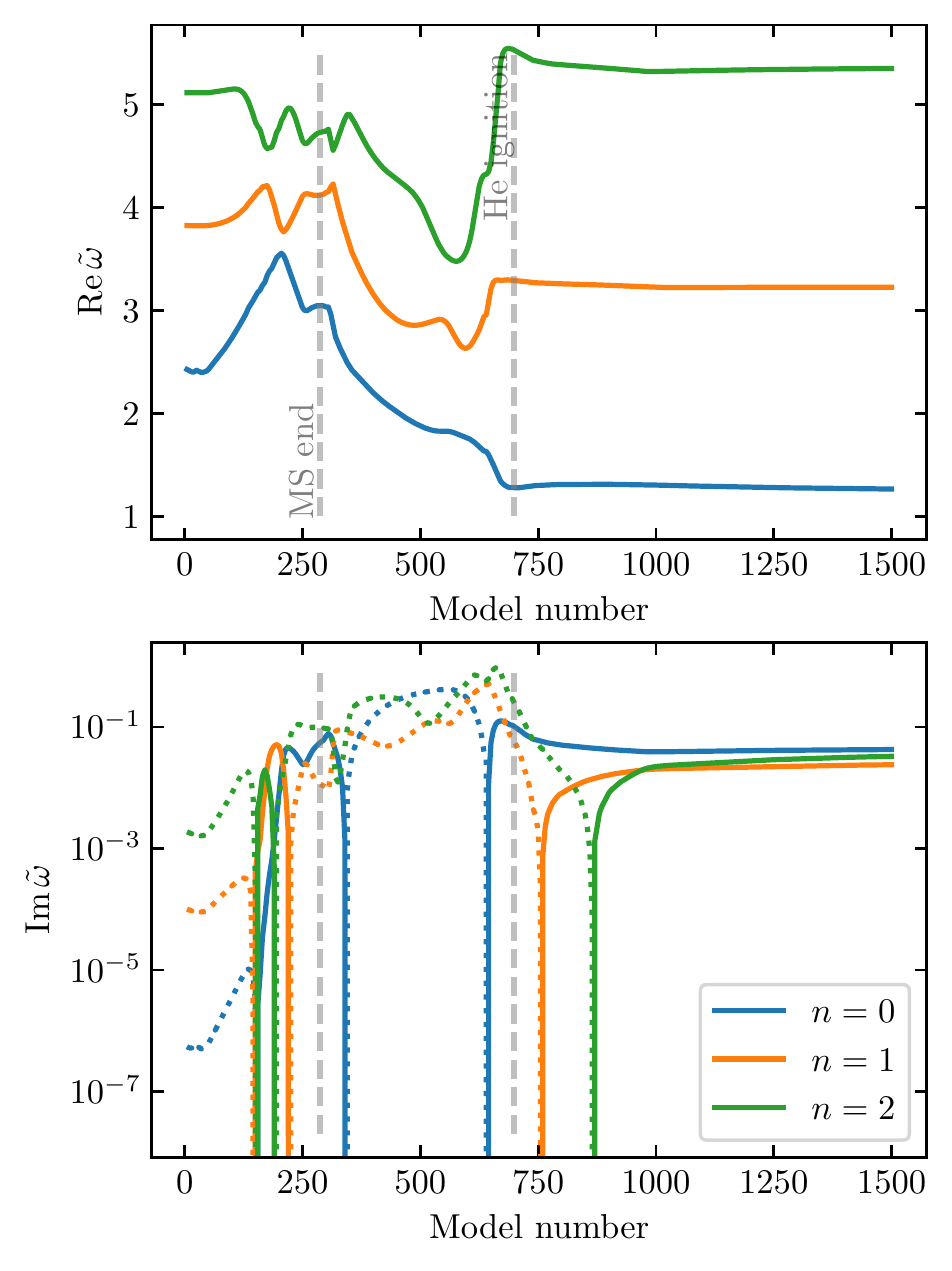}
    \caption{Same as Fig. \ref{fig:mesa-eigenvalues}, but for non-adiabatic radial modes of a non-rotating $30\,M_\odot$ star. The real parts of the oscillation frequencies ($\Re \tilde{\omega}$) are similar to those of Fig. \ref{fig:mesa-eigenvalues}. The imaginary parts ($\Im\tilde{\omega}$) are typically much larger, and much of the time the modes are damped (dashed lines) with relatively high damping rates ($|\Im \tilde{\omega}| \sim 0.1$). Hence non-adiabatic damping may overwhelm viscous driving in many cases.
    }
    \label{fig:mesa-eigenvalue-gyre-nad}
\end{figure}

For reference, we show non-adiabatic mode frequencies of a non-rotating $30 \, M_\odot$ star in Fig. \ref{fig:mesa-eigenvalue-gyre-nad}, which are calculated with the stellar pulsation code GYRE \cite[]{TownsendTeitler2013,GoldsteinTownsend2020}. The real parts of the eigenvalues are close to the adiabatic case of the rotating star (Fig. \ref{fig:mesa-eigenvalues}), while the imaginary parts are very different due to the different source of driving and damping. In the non-adiabatic case, the modes are mostly damped rather than driven, though there are some moments of instability on the main sequence and during helium burning. When compared with the viscously driven modes shown in Fig. \ref{fig:mesa-eigenvalues}, we find the non-adiabatic damping rates are higher ($|\Im \tilde{\omega}|\sim 0.1$) than the viscous driving rates ($|\Im \tilde{\omega}|\sim 10^{-5}$)  which means that non-adiabatic damping may overshadow the destabilization due to the viscous instability. However, since the growth rate of viscously driven modes is roughly proportional to the effective viscosity (see Eq. \ref{equ:low-viscosity-perturbation}), the viscous instability could operate if the effective viscosity is relatively large compared to the ad-hoc value we used in our models. Future work should attempt to calculate realistic effective viscosities to better estimate viscous driving rates.

While our study has focused on radial modes, it is possible that non-radial modes could also be driven by viscous instability. For instance, axisymmetric $\ell=2$ modes also perturb the star's moment of inertia, possibly allowing outer layers to be accelerated outwards via AM transport from inner layers during the pulsation cycle. Investigating the possibility of non-radial viscous mode excitation, including non-axisymmetric modes, will be an interesting topic for future work.

\section{Conclusions}
\label{sec:conclusions}


We have examined the impact of viscous angular momentum transport on the radial pulsations of stars. We first derived the system of equations including viscous torques, and then solved these equations for simple polytropic models. We showed that viscosity can drive the growth of radial pulsations, especially for steep outwardly decreasing rotation profiles, as expected to occur in real stars.

\begin{enumerate}
    \item When the viscosity of the star is high and the rotational profile is flat, the {oscillation frequencies of high-order modes can be fairly well approximated by fixing $f_\Omega \equiv (\delta \Omega/\Omega)/(\delta r/r)=0$ because its value averages close to zero due to global angular momentum conservation.}
    \item The contrary case of low viscosity can be well approximated by fixing $f_\Omega=-2$, which is also naturally required by local angular momentum conservation. In the low-viscosity limit, the mode growth rates can be easily computed via a work integral based on the $f_\Omega=-2$ solution. Since the effective viscosity is typically expected to be small in realistic stars, this approximation can be extended to predict when viscous instabilities occur.
    \item Applying these calculations to a $30 \, M_\odot$ stellar model with an ad-hoc viscosity profile, we find viscously driven overstable modes through much of its life time until helium burning. We do not find unstable centrifugally driven modes as speculated by \cite{ZhaoFuller2020}. The growth rate of viscously driven modes is largest near the Hertzprung gap where the star has significant differential rotation and rapid surface rotation. Hence, viscous instability may be most likely to drive pulsations and/or mass loss during this phase of stellar evolution.
\end{enumerate}

Our models do not include detailed treatments of the source of viscous angular momentum transport, so this should be improved in future work. Moreover, we focused on radial and adiabatic modes, but non-radial modes could also be driven by viscous instability, while non-adiabatic effects may often dominate mode growth/damping rates. We leave these uncertainties for future studies.

\section*{Acknowledgements}

J.F. is thankful for support through an Innovator Grant from The Rose Hills Foundation, and the Sloan Foundation through grant FG-2018-10515.

\section*{Data Availability}
Scripts involved in this work are available upon reasonable request to the authors.



\bibliographystyle{mnras}
\bibliography{bib,CoreRotBib} 

\begin{thebibliography}{}
\makeatletter
\relax
\def\mn@urlcharsother{\let\do\@makeother \do\$\do\&\do\#\do\^\do\_\do\%\do\~}
\def\mn@doi{\begingroup\mn@urlcharsother \@ifnextchar [ {\mn@doi@}
  {\mn@doi@[]}}
\def\mn@doi@[#1]#2{\def\@tempa{#1}\ifx\@tempa\@empty \href
  {http://dx.doi.org/#2} {doi:#2}\else \href {http://dx.doi.org/#2} {#1}\fi
  \endgroup}
\def\mn@eprint#1#2{\mn@eprint@#1:#2::\@nil}
\def\mn@eprint@arXiv#1{\href {http://arxiv.org/abs/#1} {{\tt arXiv:#1}}}
\def\mn@eprint@dblp#1{\href {http://dblp.uni-trier.de/rec/bibtex/#1.xml}
  {dblp:#1}}
\def\mn@eprint@#1:#2:#3:#4\@nil{\def\@tempa {#1}\def\@tempb {#2}\def\@tempc
  {#3}\ifx \@tempc \@empty \let \@tempc \@tempb \let \@tempb \@tempa \fi \ifx
  \@tempb \@empty \def\@tempb {arXiv}\fi \@ifundefined
  {mn@eprint@\@tempb}{\@tempb:\@tempc}{\expandafter \expandafter \csname
  mn@eprint@\@tempb\endcsname \expandafter{\@tempc}}}

\bibitem[\protect\citeauthoryear{{Beck} et~al.,}{{Beck} et~al.}{2012}]{beck:12}
{Beck} P.~G.,  et~al., 2012, \mn@doi [\nat] {10.1038/nature10612}, \href
  {http://adsabs.harvard.edu/abs/2012Natur.481...55B} {481, 55}

\bibitem[\protect\citeauthoryear{{Davidson}}{{Davidson}}{2020}]{davidson:20}
{Davidson} K.,  2020, \mn@doi [Galaxies] {10.3390/galaxies8010010}, \href
  {https://ui.adsabs.harvard.edu/abs/2020Galax...8...10D} {8, 10}

\bibitem[\protect\citeauthoryear{{Dziembowski}}{{Dziembowski}}{1971}]{Dziembowski1971}
{Dziembowski} W.~A.,  1971, \actaa, \href
  {https://ui.adsabs.harvard.edu/abs/1971AcA....21..289D} {21, 289}

\bibitem[\protect\citeauthoryear{{Ekstr{\"o}m}, {Meynet}, {Maeder}  \&
  {Barblan}}{{Ekstr{\"o}m} et~al.}{2008}]{ekstrom:08}
{Ekstr{\"o}m} S.,  {Meynet} G.,  {Maeder} A.,   {Barblan} F.,  2008, \mn@doi
  [\aap] {10.1051/0004-6361:20078095}, \href
  {https://ui.adsabs.harvard.edu/abs/2008A&A...478..467E} {478, 467}

\bibitem[\protect\citeauthoryear{{Fuller}, {Piro}  \& {Jermyn}}{{Fuller}
  et~al.}{2019}]{FullerPiro2019}
{Fuller} J.,  {Piro} A.~L.,   {Jermyn} A.~S.,  2019, \mn@doi [\mnras]
  {10.1093/mnras/stz514}, \href
  {https://ui.adsabs.harvard.edu/abs/2019MNRAS.485.3661F} {485, 3661}

\bibitem[\protect\citeauthoryear{{Gagnier}, {Rieutord}, {Charbonnel}, {Putigny}
   \& {Espinosa Lara}}{{Gagnier} et~al.}{2019}]{gagnier:19}
{Gagnier} D.,  {Rieutord} M.,  {Charbonnel} C.,  {Putigny} B.,   {Espinosa
  Lara} F.,  2019, \mn@doi [\aap] {10.1051/0004-6361/201834599}, \href
  {https://ui.adsabs.harvard.edu/abs/2019A&A...625A..88G} {625, A88}

\bibitem[\protect\citeauthoryear{{Glatzel}}{{Glatzel}}{1994}]{glatzel:94}
{Glatzel} W.,  1994, \mn@doi [\mnras] {10.1093/mnras/271.1.66}, \href
  {https://ui.adsabs.harvard.edu/abs/1994MNRAS.271...66G} {271, 66}

\bibitem[\protect\citeauthoryear{{Goldstein} \& {Townsend}}{{Goldstein} \&
  {Townsend}}{2020}]{GoldsteinTownsend2020}
{Goldstein} J.,  {Townsend} R.~H.~D.,  2020, \mn@doi [\apj]
  {10.3847/1538-4357/aba748}, \href
  {https://ui.adsabs.harvard.edu/abs/2020ApJ...899..116G} {899, 116}

\bibitem[\protect\citeauthoryear{{Groh}, {Hillier}  \& {Damineli}}{{Groh}
  et~al.}{2006}]{groh:06}
{Groh} J.~H.,  {Hillier} D.~J.,   {Damineli} A.,  2006, \mn@doi [\apjl]
  {10.1086/500928}, \href
  {https://ui.adsabs.harvard.edu/abs/2006ApJ...638L..33G} {638, L33}

\bibitem[\protect\citeauthoryear{{Groh} et~al.,}{{Groh} et~al.}{2009}]{groh:09}
{Groh} J.~H.,  et~al., 2009, \mn@doi [\apjl] {10.1088/0004-637X/705/1/L25},
  \href {https://ui.adsabs.harvard.edu/abs/2009ApJ...705L..25G} {705, L25}

\bibitem[\protect\citeauthoryear{{Hastings}, {Wang}  \& {Langer}}{{Hastings}
  et~al.}{2020}]{HastingsWang2020}
{Hastings} B.,  {Wang} C.,   {Langer} N.,  2020, \mn@doi [\aap]
  {10.1051/0004-6361/201937018}, \href
  {https://ui.adsabs.harvard.edu/abs/2020A&A...633A.165H} {633, A165}

\bibitem[\protect\citeauthoryear{{Jiang}, {Cantiello}, {Bildsten}, {Quataert},
  {Blaes}  \& {Stone}}{{Jiang} et~al.}{2018}]{jiang:18}
{Jiang} Y.-F.,  {Cantiello} M.,  {Bildsten} L.,  {Quataert} E.,  {Blaes} O.,
  {Stone} J.,  2018, \mn@doi [\nat] {10.1038/s41586-018-0525-0}, \href
  {https://ui.adsabs.harvard.edu/abs/2018Natur.561..498J} {561, 498}

\bibitem[\protect\citeauthoryear{{Lane}}{{Lane}}{1870}]{Lane1870}
{Lane} H.~J.,  1870, \mn@doi [American Journal of Science]
  {10.2475/ajs.s2-50.148.57}, \href
  {https://ui.adsabs.harvard.edu/abs/1870AmJS...50...57L} {50, 57}

\bibitem[\protect\citeauthoryear{{Langer}}{{Langer}}{1998}]{langer:98}
{Langer} N.,  1998, \aap, \href
  {https://ui.adsabs.harvard.edu/abs/1998A&A...329..551L} {329, 551}

\bibitem[\protect\citeauthoryear{{Papaloizou}, {Alberts}, {Pringle}  \&
  {Savonije}}{{Papaloizou} et~al.}{1997a}]{papaloizou:97}
{Papaloizou} J.~C.~B.,  {Alberts} F.,  {Pringle} J.~E.,   {Savonije} G.~J.,
  1997a, \mn@doi [\mnras] {10.1093/mnras/284.4.821}, \href
  {https://ui.adsabs.harvard.edu/abs/1997MNRAS.284..821P} {284, 821}

\bibitem[\protect\citeauthoryear{{Papaloizou}, {Alberts}, {Pringle}  \&
  {Savonije}}{{Papaloizou} et~al.}{1997b}]{saio:97}
{Papaloizou} J.~C.~B.,  {Alberts} F.,  {Pringle} J.~E.,   {Savonije} G.~J.,
  1997b, \mn@doi [\mnras] {10.1093/mnras/284.4.821}, \href
  {https://ui.adsabs.harvard.edu/abs/1997MNRAS.284..821P} {284, 821}

\bibitem[\protect\citeauthoryear{{Paxton}, {Bildsten}, {Dotter}, {Herwig},
  {Lesaffre}  \& {Timmes}}{{Paxton} et~al.}{2011}]{paxton:11}
{Paxton} B.,  {Bildsten} L.,  {Dotter} A.,  {Herwig} F.,  {Lesaffre} P.,
  {Timmes} F.,  2011, \mn@doi [\apjs] {10.1088/0067-0049/192/1/3}, \href
  {http://adsabs.harvard.edu/abs/2011ApJS..192....3P} {192, 3}

\bibitem[\protect\citeauthoryear{{Paxton} et~al.,}{{Paxton}
  et~al.}{2013}]{paxton:13}
{Paxton} B.,  et~al., 2013, \mn@doi [\apjs] {10.1088/0067-0049/208/1/4}, \href
  {http://adsabs.harvard.edu/abs/2013ApJS..208....4P} {208, 4}

\bibitem[\protect\citeauthoryear{{Paxton} et~al.,}{{Paxton}
  et~al.}{2015}]{paxton:15}
{Paxton} B.,  et~al., 2015, \mn@doi [\apjs] {10.1088/0067-0049/220/1/15}, \href
  {http://adsabs.harvard.edu/abs/2015ApJS..220...15P} {220, 15}

\bibitem[\protect\citeauthoryear{{Paxton} et~al.,}{{Paxton}
  et~al.}{2018}]{paxton:18}
{Paxton} B.,  et~al., 2018, \mn@doi [\apjs] {10.3847/1538-4365/aaa5a8}, \href
  {http://adsabs.harvard.edu/abs/2018ApJS..234...34P} {234, 34}

\bibitem[\protect\citeauthoryear{{Paxton} et~al.,}{{Paxton}
  et~al.}{2019}]{PaxtonSmolec2019}
{Paxton} B.,  et~al., 2019, \mn@doi [\apjs] {10.3847/1538-4365/ab2241}, \href
  {https://ui.adsabs.harvard.edu/abs/2019ApJS..243...10P} {243, 10}

\bibitem[\protect\citeauthoryear{{Sana} et~al.,}{{Sana} et~al.}{2012}]{sana:12}
{Sana} H.,  et~al., 2012, \mn@doi [Science] {10.1126/science.1223344}, \href
  {http://adsabs.harvard.edu/abs/2012Sci...337..444S} {337, 444}

\bibitem[\protect\citeauthoryear{{Smith}}{{Smith}}{2014}]{smithrev:14}
{Smith} N.,  2014, \mn@doi [\araa] {10.1146/annurev-astro-081913-040025}, \href
  {https://ui.adsabs.harvard.edu/abs/2014ARA&A..52..487S} {52, 487}

\bibitem[\protect\citeauthoryear{{Smith}}{{Smith}}{2017}]{smith:17}
{Smith} N.,  2017, \mn@doi [Philosophical Transactions of the Royal Society of
  London Series A] {10.1098/rsta.2016.0268}, \href
  {https://ui.adsabs.harvard.edu/abs/2017RSPTA.37560268S} {375, 20160268}

\bibitem[\protect\citeauthoryear{{Spruit}}{{Spruit}}{2002}]{spruit:02}
{Spruit} H.~C.,  2002, \mn@doi [\aap] {10.1051/0004-6361:20011465}, \href
  {http://adsabs.harvard.edu/abs/2002A%26A...381..923S} {381, 923}

\bibitem[\protect\citeauthoryear{{Townsend} \& {Teitler}}{{Townsend} \&
  {Teitler}}{2013}]{TownsendTeitler2013}
{Townsend} R.~H.~D.,  {Teitler} S.~A.,  2013, \mn@doi [\mnras]
  {10.1093/mnras/stt1533}, \href
  {https://ui.adsabs.harvard.edu/abs/2013MNRAS.435.3406T} {435, 3406}

\bibitem[\protect\citeauthoryear{{Unno}, {Osaki}, {Ando}, {Saio}  \&
  {Shibahashi}}{{Unno} et~al.}{1989}]{unno:89}
{Unno} W.,  {Osaki} Y.,  {Ando} H.,  {Saio} H.,   {Shibahashi} H.,  1989,
  {Nonradial oscillations of stars}.
University of Tokyo Press, Tokyo

\bibitem[\protect\citeauthoryear{{Vink}}{{Vink}}{2021}]{vink:21}
{Vink} J.~S.,  2021, arXiv e-prints, \href
  {https://ui.adsabs.harvard.edu/abs/2021arXiv210908164V} {p. arXiv:2109.08164}

\bibitem[\protect\citeauthoryear{{Wu} \& {Fuller}}{{Wu} \&
  {Fuller}}{2021}]{wu:21}
{Wu} S.,  {Fuller} J.,  2021, \mn@doi [\apj] {10.3847/1538-4357/abc87c}, \href
  {https://ui.adsabs.harvard.edu/abs/2021ApJ...906....3W} {906, 3}

\bibitem[\protect\citeauthoryear{{Yadav} \& {Glatzel}}{{Yadav} \&
  {Glatzel}}{2017}]{yadav:17}
{Yadav} A.~P.,  {Glatzel} W.,  2017, \mn@doi [\mnras] {10.1093/mnras/stx1808},
  \href {https://ui.adsabs.harvard.edu/abs/2017MNRAS.471.3245Y} {471, 3245}

\bibitem[\protect\citeauthoryear{{Zhao} \& {Fuller}}{{Zhao} \&
  {Fuller}}{2020}]{ZhaoFuller2020}
{Zhao} X.,  {Fuller} J.,  2020, \mn@doi [\mnras] {10.1093/mnras/staa1097},
  \href {https://ui.adsabs.harvard.edu/abs/2020MNRAS.495..249Z} {495, 249}

\makeatother
\end{thebibliography}




{

\appendix

\section{Derivation with Eulerian perturbations}
\label{sec:eulerian_perturbations}

To demonstrate the validity of Eq. \eqref{equ:euler-radial} and \eqref{equ:euler-horizontal}, we show that they are equivalent to those derived from Eulerian perturbations. 
We start with Eq. (32.17) of \cite{unno:89}, which treated Eulerian perturbations of a rotating star {in an inertial frame. The rotation rate $\Omega$ is defined via ${\bf v}_0 = \mathbf{\Omega} \times {\bf r} = \Omega r \sin \theta\hat{\phi}$, where ${\bf v}_0$ is the rotational velocity in the inertial frame.} Note that when operating on a scalar quantity, $\dd /\dd t = \partial/\partial t + (\vec{v} \cdot \nabla ) = \partial/\partial t + \Omega \partial/\partial\phi$. Hence we can write Eq. (32.17) of \cite{unno:89} as
\begin{align}
    \frac{\dd v'_i}{\dd t} \hat{e}_i+ 2 \vec{\Omega} \times \vec{v}' + (\vec{v}'\cdot \nabla \Omega) r \sin \theta \hat{\phi} \nonumber \\= -\frac{\nabla p'}{\rho} -\nabla \Phi' + \frac{\rho'}{\rho^2} \nabla p +  \vec{f}'_{\rm vis}. \label{equ:unno32.17}
\end{align}
Here $\hat{e}_i = \{\hat{r}, \hat{\theta}, \hat{\phi}\}$ are the spherical coordinate unit vectors, and the first term has adopted the Einstein summation convention. Eulerian perturbations are labeled with primes. We also added a term $\vec{f}'_{\rm vis}$, which is the perturbation of the viscous force and it only has a $\hat{\phi}$ component. Caution must be taken when taking time derivatives, because the unit vectors $\hat{e}_i$ have non-zero total time derivatives.

Next we relate the Eulerian and Lagrangian perturbations. Eq. (32.18) of \cite{unno:89} gives the relation between $\vec{v}'$ and the Lagrangian displacement $\vec{\xi}$ (which is equivalent to $\delta \vec{r}$ in the main text),
\begin{align}
    \vec{v}' = \frac{\dd (\xi_i)}{\dd t} \vec{e}_i - (\vec{\xi}\cdot \nabla \Omega) r\sin \theta \hat{\phi}.
\end{align}
Note that $\hat{\phi} \cdot \nabla \vec{\Omega}=0$. {Since we only study radial modes, axisymmetry is assumed and thus $\phi$-derivatives are ignored.} After some algebra, the left hand side of Eq. \eqref{equ:unno32.17}  becomes
\begin{align}
    \mathrm{LHS} =  \frac{\partial^2}{\partial t^2} \vec{\xi} + 2 \vec{\Omega} \times \frac{\partial}{\partial t} \vec{\xi} + ( \vec{\xi} \cdot\nabla \Omega^2 ) \big( r \sin^2 \theta \hat r + r \sin \theta \cos \theta \hat\theta \big).\label{equ:lhs}
\end{align}

We then relate the Eulerian and Lagrangian perturbations in $\rho$, $p$, and $\Phi$, they both follow the general form like $\delta \rho = \rho'+ \vec{\xi}\cdot \nabla \rho $. Moreover, there are additional relations:
\begin{itemize}
    \item adiabatic equation of state: $\delta p/p = \Gamma_1 \delta \rho/\rho$;
    \item force equilibrium: $ \nabla p = - \rho g \hat{r} + \rho \Omega^2 r (\sin \theta \hat{r} + \cos \theta \hat{\theta})$;
    \item mass conservation: Eq. \eqref{equ:mass-conservation};
    \item gravity perturbation: {$\dd \Phi'/\dd r = g' = -4\pi G \rho \xi_r$}.
\end{itemize}
After copious algebra, {the right hand side} of equation \ref{equ:unno32.17} becomes
\begin{align}
\label{equ:rhs}
    \mathrm{RHS} & = \left[\frac{-p}{\rho} \frac{\partial}{\partial r} \bigg(\frac{\delta p}{p}\bigg) - \frac{1}{\rho} \frac{\partial p}{\partial r} \bigg( \frac{\delta p}{p} + 2 \frac{\xi_r}{r} \bigg) + 2 g \frac{\xi_r}{r} + \Omega^2 r \sin^2 \theta \frac{\xi_r}{r}\right.\nonumber\\ & \left.+ r^2 \sin^2 \theta \frac{\partial \Omega^2}{\partial r} \frac{\xi_r}{r} \right] \hat{r} + f'_{\rm vis} \hat{\phi}.
\end{align}

Combining the $\hat{r}$ components of the LHS and the RHS recovers Eq. \eqref{equ:euler-radial}, after integrating over $\theta$ and redefining $\Omega$ as we do in the main text. We have also used the definition $\partial {\xi}_\phi/\partial t \equiv r\delta \Omega \sin\theta$. Note that the differential rotation term in equation \ref{equ:rhs} arises from taking the radial derivative of the background pressure gradient, and it cancels the $r$-component of the differential rotation term on the left-hand side of equation \ref{equ:lhs}.
The $\hat \phi$ component of equation \ref{equ:rhs} is just ${f}'_{\rm vis}=\delta {f}_{\rm vis}-(\vec\xi \cdot \nabla){f}_{\rm vis}$. The $\hat{\phi}$ component of the LHS is $\ddot{\xi}_\phi + 2 \Omega \dot \xi_r \sin \theta $. Combining the LHS and RHS recovers Eq. \eqref{equ:euler-horizontal}, so the $\phi-$component of our equation of motion is also verified.


{Above, we equated the time-derivative of $\xi_\phi$ with a ``perturbed" rotation rate $\delta \Omega$. This is equivalent to using an accelerating reference frame that remains co-rotating with the shell during the pulsation cycle, as we do in the main text. In this accelerating frame, $\xi_\phi = 0$ because we only consider radial modes, but $d \Omega/dt$ is non-zero. The Euler force becomes
\begin{align}
    \frac{d \vec{\Omega}}{dt} \times \vec{r} &= \frac{d \delta \vec{\Omega}}{dt} \times \vec{r} \nonumber \\
    &= \delta \dot{\Omega} \, \hat{z} \times \vec{r} \nonumber \\
    &= \delta \dot{\Omega} \, r \sin \theta \hat{\phi} \, .
\end{align}
This is identical to the term $\partial^2 \xi_\phi/\partial t^2 = (\partial/\partial t) (\delta \Omega r \sin \theta) = \delta \dot{\Omega} r \sin \theta$ that appears in the analysis above in a frame with fixed rotation rate, which accounts for the first term in equation \ref{equ:euler-horizontal}. So regardless of whether one adopts an inertial frame, a frame with fixed rotation rate, or a co-rotating but accelerating frame, the eigenvalue equations are identical, as they should be.}

}

\section{Solving perturbation equations}
\label{app:solving-perturbation-equations}

Here we describe how we solved the complex differential equations \ref{equ:4-parameter-equations-y1}-\ref{equ:4-parameter-equations-y4} including viscosity. The equations and variables are both dimensionless, in the form of
\begin{align}
    x \frac{\dd \Vec{y}}{\dd x} = \mathbf{A}(\omega,x) \Vec{y}. \label{equ:univeral-form}
\end{align}
where $\mathbf{A}$ is an $n\times n$ matrix, $n$ is the number of variables (four in this case), and $\vec{y}$ is a vector consisting of the $n$ variables. There are also inner and outer boundaries
\begin{align}
    \mathbf{B}_{\rm i} (x_1) \Vec{y}_1 =0, \qquad \mathbf{B}_{\rm o} (x_N) \vec{y}_N = 0.\label{equ:bc_discretized}
\end{align}
Here $\mathbf{B}_{\rm i}$ and $\mathbf{B}_{\rm o}$ are matrices with $n$ columns; $x_1$ and $x_N$ are endpoints (while the subscripts indicate the discretized grid point, assuming there are $N$ grid points).

We may also include integration relations, in the form of
\begin{align}
    \int_{x_1}^{x_N} \mathbf{I}(x) \vec{y} \dd x = 0.\label{equ:universal-int}
\end{align}
Here $\mathbf{I}$ is a matrix with $n$ columns. The total number of boundary conditions and integration conditions should be $n$.

\subsection{Discretizing the equations}
We discretize $x$ into $N-1$ divisions and label them as $x_1, \dots, x_N$. For the $j$-th point, we may use the differential stencil so that Eq. \eqref{equ:univeral-form} becomes
\begin{align}
    \frac{x_j+x_{j-1}}{2} \frac{\Vec{y}_{j} - \Vec{y}_{j-1}}{x_j - x_{j-1}} \doteq \mathbf{A}\left(\omega,\frac{x_j+x_{j-1}}{2} \right) \frac{\vec{y}_j+\vec{y}_{j-1}}{2}.\label{equ:equ_discritized}
\end{align}
This means that $\vec{y}_j$ and $\vec{y}_{j-1}$ can be related by $\mathbf{L}_j \vec{y}_{j-1} + \mathbf{R}_j \vec{y}_{j}=0$, where $\mathbf{L}_j$ and $\mathbf{R}_j$ are both $n\times n$ matrices.

Eq. \eqref{equ:universal-int} can be approximated with the trapezoidal integration scheme as
\begin{align}
    \sum_{j=2}^{N} \mathbf{I}\left(\frac{x_j+x_{j-1}}{2}\right) \frac{\vec{y}_j+\vec{y}_{j-1}}{2} (x_j-x_{j-1}) \doteq 0.\label{equ:int_discretized}
\end{align}
Equivalently, the trapezoidal integration scheme can also be written as $\sum_{j=1}^{N} w_j \vec{y}_j=0$, where $w_j$ is the coefficient corresponding to $y_j$. Thus $(w_1,w_2,\cdots,w_N)^{\rm T}$ is also a row vector.

If we assume $\vec{Y} = (\vec{y}_1,\dots,\vec{y}_N)$, which is a vector with $nN$ elements, then all those pieces (Eqs. \ref{equ:bc_discretized}, \ref{equ:equ_discritized}, \ref{equ:int_discretized}) can be written as a large sparse $nN\times nN$ matrix $\mathbf{S}(\omega)$ such that $\mathbf{S}\vec{Y}=0$. When $\omega$ is an eigenvalue we must have $\det \mathbf{S}(\omega)=0$, thus the next step is to search for roots to the function defined as $\mathcal{D}(\omega) = \det \mathbf{S}(\omega)$.

\subsection{Searching for eigenvalues}

Without viscous or non-adiabatic effects, the eigenvalue is constricted to be real, which means we need to solve for $\mathcal{D}(\omega)=0$ along the real $\omega$ axis. The numerical solution is not complicated.

Once the eigenvalue is not guaranteed to be real, we expect the function $\mathcal{D}(\omega)$ to be complex as well. Naively, $\mathcal{D}(\omega)=0$ means $\Re \mathcal{D}(\omega)=0$ and $\Im \mathcal{D}(\omega)=0$. When we plot contours of both $\Re \mathcal{D}(\omega)=0$ and $\Im \mathcal{D}(\omega)=0$ on the complex plane, we will see eigenvalues are at crosses of the two groups of contours \cite[also see][]{GoldsteinTownsend2020}. At each eigenvalue, we see quadrants due to signs of $\Re \mathcal{D}(\omega)$ and $\Im \mathcal{D}(\omega)$.

We briefly introduce our numerical root-finding technique below.
\begin{enumerate}
    \item Discretize a rectangular complex region in the real and imaginary components of $\omega$ into a mesh. Then we calculate $\Re \mathcal{D}(\omega)$ and $\Im \mathcal{D}(\omega)$ at all mesh points. By calculating $\abs{\mathcal{D}(\omega)}$ initial guesses for the eigenvalues can be found.
    \item Near the vicinity of an initial guess, we can also search for 4 points in different quadrants of the contours, where signs of $(\Re \mathcal{D}(\omega), \Im \mathcal{D}(\omega))$ are $(+,+)$, $(+,-)$, $(-,-)$, and $(-,+)$. Within a certain radius near the the initial guess, we may find multiple candidates in each quadrant.
    \item Find the quadrilateral (made with 4 points, each from a different quadrant) that will enclose the initial guess. This can be achieved with two criteria: i) the quadrilateral encloses the initial guess; ii) the minimum distance from the initial guess to the edges of the quadrilateral should be sufficiently large.
    \item Shrink the optimal quadrilateral found above. For example, start with the edge connecting points with signs of $(+,+)$ and $(+,-)$ and check $\mathcal{D}$ at their midpoint: if signs of $\mathcal{D}$ are the same with the $(+, +)$ one, replace the $(+, +)$ representative with the midpoint, or vise versa. 
    \item Iterate to the next neighboring edge following the procedure in the step above until convergence. The numerical approximation of $\omega$ can be given as the arithmetic mean of the 4 points, while the tolerance can be estimated from size of the quadrilateral.
\end{enumerate}
The method is robust once we find a proper initial guess and an initial quadrilateral, which means the mesh of $\omega$ can not be too sparse. Moreover, the regime used to search for the initial quadrilateral can not be too large, otherwise curvatures of the contours are possible to spoil convergence. The problem can be mitigated once we have a finer mesh in the $\omega$ plane.

\begin{figure}
    \centering
    \includegraphics[width=\linewidth]{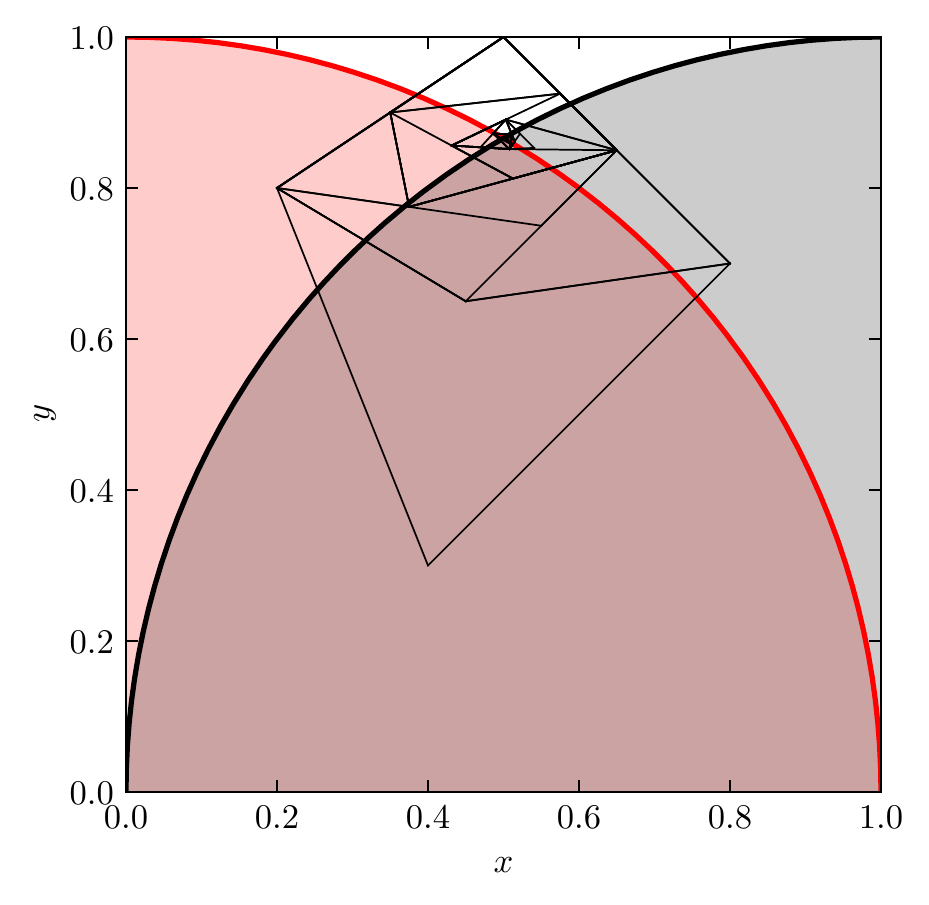}
    \caption{Example of our root finding method. Here the red (black) contour and red (black) shaded region satisfy $f(x,y)=1-x^2-y^2 \ge 0$ (or $g(x,y)=1-(x-1)^2-y^2 \ge 0$). We choose four initial points in each quadrant and then iteratively shrink the edges of the quadrilateral to converge to the solution of $f(x,y)=g(x,y)=0$. }
    \label{fig:zigzag}
\end{figure}

\begin{figure}
    \centering
    \includegraphics[width=\linewidth]{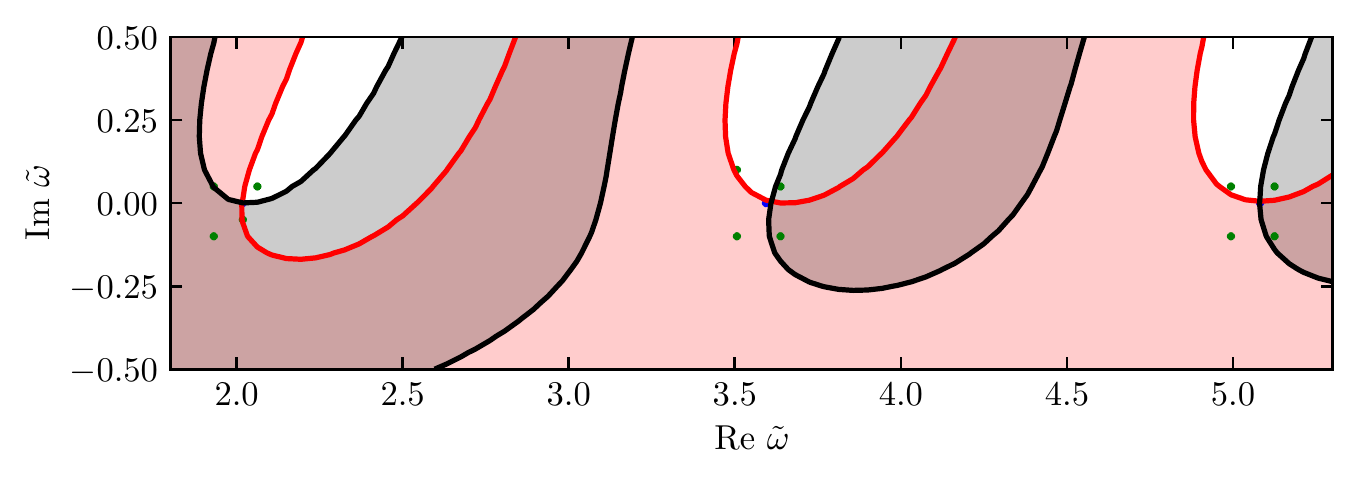}
    \includegraphics[width=\linewidth]{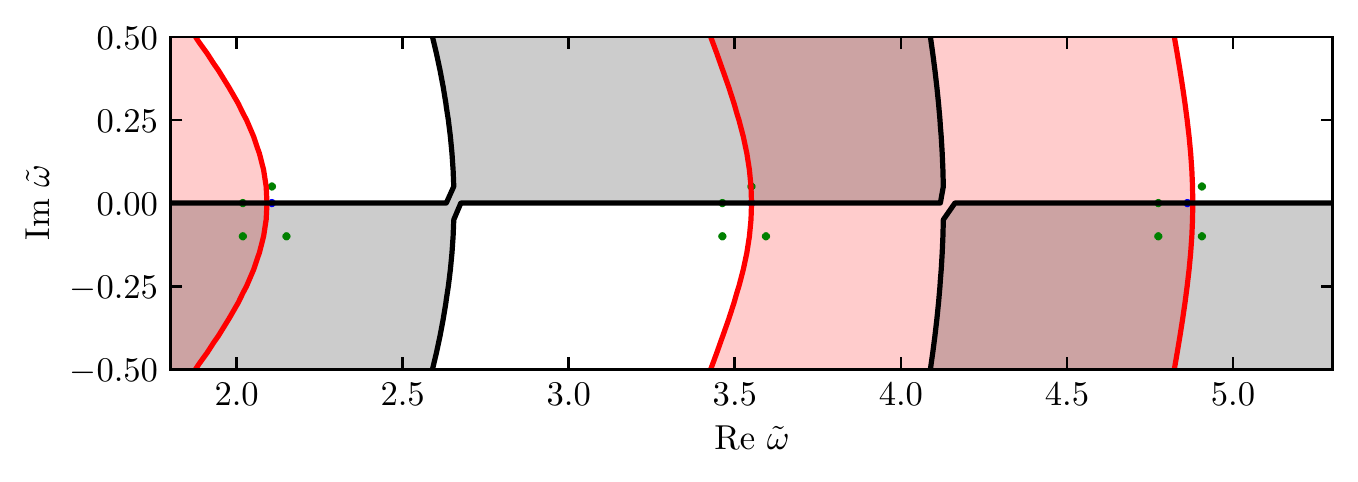}
    \caption{Contours of the determinant's real (red) and imaginary (black) parts in the complex plane, where the colored shaded regions are positive. \textbf{Top}: the low-viscosity model shown in Fig. \ref{fig:polytrope-low-viscosity}, where contours are asymmetric, and four initial points are also shown (green dots).
    \textbf{Bottom}: the high-viscosity condition, where the contours are nearly symmetric relative to $\Im \tilde{\omega}=0$, which means the eigenvalues are almost real.
    }
    \label{fig:contours}
\end{figure}

As an illustration we present an example in Fig. \ref{fig:zigzag}, where we show two contours of $1-x^2-y^2=0$ and $1-(x-1)^2-y^2=0$ in the $x-y$ plane. The cross section of the two contours where both equations are satisfied is located at $(1/2, \sqrt{3}/2)$. We first place 4 initial points near in the quadrants and iteratively shrink the edges of the quadrilateral following the routine described above, and we see convergence to the solution after several iterations. We also show examples of high- and low-viscosity polytropic models in Fig. \ref{fig:contours}.

After finding the eigenvalues, we can solve for the eigenvectors $\vec{Y}$ by solving the homogeneous linear equation for $\mathbf{S}$ or by integrating the differential equations \cite[e.g., the Magnus integrator in][]{TownsendTeitler2013}.

\section{MESA inlist}

This inlist will be uploaded to Zenodo.org upon acceptance of this paper.

\begin{Verbatim}
! centrifugal mass loss

&star_job
	pgstar_flag = .true.
	relax_initial_Z = .true.
	new_Z = 0.017d0
	new_rotation_flag = .true.
	change_rotation_flag = .true.
	new_omega_div_omega_crit = 0.25
	set_initial_omega_div_omega_crit = .true.
/ !end of star_job namelist

&controls
	use_other_wind = .false.
	!fitted_fp_ft_i_rot = .true.
	!w_div_wcrit_max = 0.7

	!-------------------------------- Convergence
	okay_to_reduce_gradT_excess = .true.
	gradT_excess_age_fraction = 0.999d0
	gradT_excess_max_change = 0.01d0
	timestep_factor_for_retries = 0.8
	timestep_factor_for_backups = 0.8
	min_timestep_factor = 0.9
	max_timestep_factor = 1.2d0
	backup_hold = 10
	retry_hold = 3
	redo_limit = -1
	relax_hard_limits_after_retry = .false.
	newton_iterations_limit = 7
	max_model_number = 30000
	max_number_retries = 5000
	! Fixing_the_position_of_the_Lagrangian_region
	! of_the_mesh_helps
	! convergence_near_the_Eddington_limit
	max_logT_for_k_below_const_q = 100
	max_q_for_k_below_const_q = 0.995
	min_q_for_k_below_const_q = 0.995
	max_logT_for_k_const_mass = 100
	max_q_for_k_const_mass = 0.99
	min_q_for_k_const_mass = 0.99
	!extra_spatial_resolution
	max_dq = 0.02
	fix_eps_grav_transition_to_grid = .true.
	! extra_controls_for_timestep
	! these_are_for_changes_in_mdot_at_the_onset_of
	! mass_transfer
	delta_lg_star_mass_limit = 1d-3
	delta_lg_star_mass_hard_limit = 2d-3
	! these_are_to_properly_resolve_core_hydrogen
	! depletion
	delta_lg_XH_cntr_limit = 0.04d0
	delta_lg_XH_cntr_max = 0.0d0
	delta_lg_XH_cntr_min = -4.0d0
	delta_lg_XH_cntr_hard_limit = 0.06d0
	! this_is_mainly_to_resolve_properly_when_the_star
	! goes_off_the_main_sequence
	delta_HR_limit = 0.002d0
	! delta_lgR_limit = 0.001d0
	! delta_lgR_hard_limit = 0.001d0
	delta_lgTeff_limit = 0.002d0
	! relax_default_dHe/He,_otherwise_growing_He_core
	! can_cause_things_to_go_at_a_snail_pace
	dHe_div_He_limit = 2.0
	! we're_not_looking_for_much_precision_at_the
	! very_late_stages
	dX_nuc_drop_limit = 5d-2

	! !-------------------------------- Rotation
	am_nu_ST_factor = 0
	smooth_nu_ST = 5
	smooth_D_ST = 5
	use_other_am_mixing = .true.
	am_time_average = .true.
	premix_omega = .true.
	recalc_mixing_info_each_substep =.true.
	am_nu_factor = 1
	am_nu_non_rotation_factor = 1d0
	am_nu_visc_factor = 1
	am_nu_ES_factor = 1
	angsml = 0.0
	am_D_mix_factor = 3.33d-2
	D_ES_factor = 1
	! this_is_to_avoid_odd_behaviour_when_a_star
	! switches_from_accreting_to_mass_losing
	max_mdot_jump_for_rotation = 1d99

	!------------------------------------ MAIN
	initial_mass = 30
	initial_z = 0.02
	use_Type2_opacities = .true.
	Zbase = 0.017d0
	predictive_mix(1) = .true.
	predictive_superad_thresh(1) = 0.005
	predictive_avoid_reversal(1) = 'he4'
	predictive_zone_type(1) = 'any'
	predictive_zone_loc(1) = 'core'
	predictive_bdy_loc(1) = 'top'
	dX_div_X_limit_min_X = 1d-4
	dX_div_X_limit = 5d-2
	dX_nuc_drop_min_X_limit = 1d-4
	dX_nuc_drop_limit = 5d-2

	!------------------------------------ WIND
	hot_wind_scheme = 'Dutch'
	cool_wind_RGB_scheme = 'Dutch'
	cool_wind_AGB_scheme = 'Dutch'
	RGB_to_AGB_wind_switch = 1d-4
	Dutch_scaling_factor = 0.1 !0.5
	mdot_omega_power = 0.43		!

	! !---------------------------- OVERSHOOTING
	 overshoot_f_above_nonburn_core = 0.02
	 overshoot_f0_above_nonburn_core = 0.005
	 overshoot_f_above_nonburn_shell = 0.02
	 overshoot_f0_above_nonburn_shell = 0.005
	 overshoot_f_below_nonburn_shell = 0.02
	 overshoot_f0_below_nonburn_shell = 0.005
	 overshoot_f_above_burn_h_core = 0.02
	 overshoot_f0_above_burn_h_core = 0.005
	 overshoot_f_above_burn_h_shell = 0.02
	 overshoot_f0_above_burn_h_shell = 0.005
	 overshoot_f_below_burn_h_shell = 0.02
	 overshoot_f0_below_burn_h_shell = 0.005
	 set_min_D_mix = .true.
	 min_D_mix = 1d2

	!------------------------------------- MISC
	photo_interval = 10
	profile_interval = 5
	max_num_profile_models = 3000
	history_interval = 1
	terminal_interval = 10
	write_header_frequency = 10
	max_number_backups = 500
	relax_max_number_retries = 2000
	max_number_retries = 4000

	!------------------------------------- MESH
	mesh_delta_coeff = 0.7
	varcontrol_target = 3d-4

	!------------------------------------- GYRE
	write_pulse_data_with_profile = .true.
	pulse_data_format = 'GYRE'
	! set_uniform_am_nu_non_rot = .true.
	! uniform_am_nu_non_rot = 1d50
/ ! end of controls namelist

&pgstar                                                                                                                                                                                                                                                                                                                                                                                                                               read_extra_pgstar_inlist1 = .true.                                                                                                                                                                               extra_pgstar_inlist1_name = 'inlist_pgstar'                                                                                                                                                                                                                                                                                                                                                                                   / ! end of pgstar namelist
\end{Verbatim}


\bsp	
\label{lastpage}
\end{document}